 \documentclass[final,5p,times,twocolumn]{elsarticle}
 \usepackage{graphicx}

\usepackage{amssymb}

\journal{Nuclear Instruments and Methods in Physics Research A}

\begin{document}

\begin{frontmatter}


\title{Simulation and Analysis of TE Wave Propagation for Measurement of Electron Cloud Densities in Particle Accelerators}

\author[label1]{Kiran G. Sonnad}
\author[label3]{Kenneth Hammond}
\author[label1]{Robert Schwartz}
\author[label4]{Seth A. Veitzer}
\address[label1]{CLASSE, Cornell University, Ithaca NY, USA}
\address[label3]{Department of Physics, Harvard University, Cambridge MA, USA }
\address[label4]{Tech-X Corporation, Boulder CO, USA}

\begin{abstract}
The use of transverse electric (TE) waves has proved to be a powerful, noninvasive method for
estimating the densities of electron clouds formed in particle accelerators. Results from 
the plasma simulation program VSim have served as a useful guide for experimental studies related 
to this method, which have been performed at various accelerator facilities. This paper provides 
results of the simulation and modeling work done in conjunction with experimental efforts carried 
out at the Cornell electron storage ring ``Test Accelerator" (CesrTA). This paper begins with a 
discussion of the phase shift induced by electron clouds in the transmission of RF waves, followed by 
the effect of reflections along the beam pipe, simulation of the resonant standing wave frequency 
shifts and finally the effects of external magnetic fields, namely dipoles and wigglers. A derivation 
of the dispersion relationship of wave propagation for arbitrary geometries in field free regions with 
a cold, uniform cloud density is also provided.    
\end{abstract}


\end{frontmatter}


\section{Introduction}
\label{}
Electron clouds formed in circular particle accelerators with positively 
charged beams are known to degrade the quality of the beam. They are a 
concern for future accelerator facilities such as the International Linear
Collider (ILC) damping rings, the SuperKEKB, and also upgrade of existing
facilities such as the Large Hadron Collider (LHC) and the Fermilab Main
Injector (MI). Their study is also important for the optimum performance
of spallation neutron sources.

The detection of electron clouds has been a topic of study ever since
their effects were first observed.  The methods used for such a detection
have included retarding field analyzers, clearing electrodes, shielded
pickup detectors, TE waves and study of the response of the beam in the 
presence of an electron cloud. The TE wave method involves transmitting 
microwaves through a section of the beam pipe, and then studying the 
effect of the cloud on the microwave properties. The microwaves can be 
introduced either as traveling or standing waves within a section of the 
beam pipe.

The method of using microwaves as a probe for investigating
the presence of electron clouds was first proposed by 
F. Caspers~\cite{Caspers,Kroyer}, 
in which experiments were conducted at the SPS at
CERN based on this method. The measurement technique involves measuring
the height of modulation side-bands off the carrier frequency of the
microwave. The electron cloud, constituting a plasma modifies the
dispersion relationship of the microwave. The periodic production and
clearing of the cloud, based on the bunch train passage frequency, leads
to modulation of the phase advance of carrier wave as it travels. This 
modulation gives rise to side bands in the spectrum of the propagated wave. 
The side bands are spaced from the carrier frequency by a value equal to 
integer multiples of the train passage frequency. The details of the
distribution of all the side band heights depends upon the nature of
the build up and decay of the cloud. The study of this paper is restricted
to the effect of the wave dispersion from a static cloud under various conditions.  
A confirmation of the electron
cloud induced modulation was shown in the PEP II Low Energy Ring (LER)
\cite{StefanoPRL}. In this experiment, the wave was transmitted across
a solenoidal section of the ring and the cloud density was controlled by
adjusting the strength of the solenoidal magnetic field. The estimations
of the electron cloud density in this experiment, done based on the
formulation given in \cite{SonnadPAC}, were reasonable when compared to
earlier build up simulations.

As shown in this paper, reflections within the beam pipe can alter the signal 
and thus misrepresent the cloud density in the region being sampled. Instead 
of transmitting the wave at a point and receiving it from another called 
traveling wave RF diagnostics, one could also trap the wave within a section
called resonant wave RF diagnostics. This has the advantage that
one can sample a known section of the beam pipe. This method would not be 
affected by waves being reflected from other segments of the pipe
coming back into the section of interest and thus compromising the precision
of the measurement. Another advantage of trapping, is that there is an enhancement 
of the signal as long as the point of measurement is close to a peak of the standing wave. 
Additionally, at resonance, as demonstrated in Ref~\cite{SonnadANKA}, there is improved
matching of the signal transfer from the electrodes into the waveguide,
which is the beam pipe. As discussed in Ref~\cite{SikoraIPAC11}, the
modulation of the cloud density would result in a modulation of the
resonant frequency enabling one to relate the frequency modulation signal to an
actual cloud density. The draw back of this technique is the need for 
having reflectors at both ends of the desired section. 

The electron cloud induced phase shift is known to undergo an enhancement 
in the presence of an external magnetic field under certain conditions, 
due to a modification in the dispersion relationship. This occurs when 
the wave magnetic field has a component perpendicular to the external 
magnetic field, and the frequency of the wave is close to the electron 
cyclotron frequency, corresponding to the value of the external magnetic 
field. This effect was demonstrated through simulations \cite{SonnadAPS, VeitzerDOE}, 
and was later confirmed through experiments done at PEP II across a 
chicane \cite{PiviEPAC}. Further measurements done at the same chicane, 
now installed in CesrTA also confirm this effect. While it is good to have
an enhanced signal, the drawback of such a measurement is that there is no
available formulation that relates the enhanced side-band amplitude with
the expected electron cloud density. In addition, as discussed in
this paper, the phase shift progressively reduces as the electron cyclotron
frequency exceeds the carrier wave frequency and at very high external
magnetic fields the signal may not be observable at all. On can suppress 
the effect of the external magnetic field by aligning the wave electric 
field parallel to the external magnetic field, however it will be shown 
in this paper that this cannot be fully eliminated unless the waveguide 
is rectangular.

The program VSim\cite{Vorpal}, previously Vorpal, was used throughout to 
perform the simulations. The simulations used electromagnetic particle-in-cell (PIC) algorithms, 
consisting of propagating waves through a conducting beam pipe. The
end of the pipe had perfectly-matched layers (PMLs) [7] meant to absorb any transmissions, 
and thus simulate a long, continuous beam pipe. Electrons were uniformly distributed and set
to initially have zero velocity (a cold plasma). Waves were excited in the simulations with 
the help of a vertically pointing current density near one end of the beam pipe, which covered
the full cross-sectional area and was two cells thick in the longitudinal dimension. 
In later simulations the PMLs and RF current source were replaced with port boundary conditions
that simultaneously absorb RF energy at a single frequency at the ends of the simulation while
also launching RF energy into the simulation domain to simulate traveling waves.

The waves were propagated along the channel using the Yee [5] algorithm for solving the electromagnetic
field equations. The computational parameters used did not change much between the various
simulations performed in this paper. All the simulations were three dimensional using a
Cartesian grid. The grid sizes were around $2-3 \mbox{mm}$ in each direction, the time steps varied
between $2-4 \times 10^{-12}s$. The macro-particles-per-cell used was typically 10, and they
were loaded uniformly in position space, with zero initial velocity, often referred to as 
a ``cold start". The duration of the simulation was about 140 RF cycles.    

Overall, the modeling effort related to measuring electron clouds using
TE waves has served as a useful guide toward better understanding of the 
physical phenomenon and proper interpretation of the measured data. This 
paper provides a comprehensive account of simulations performed for various
techniques currently under study. A derivation of the wave dispersion
relationship for propagation through a beam pipe with a cold, uniform electron 
distribution in a field free region is given in the Appendix. 
\begin{figure}[ht]
\begin{center}
\includegraphics*[width=0.5\columnwidth]{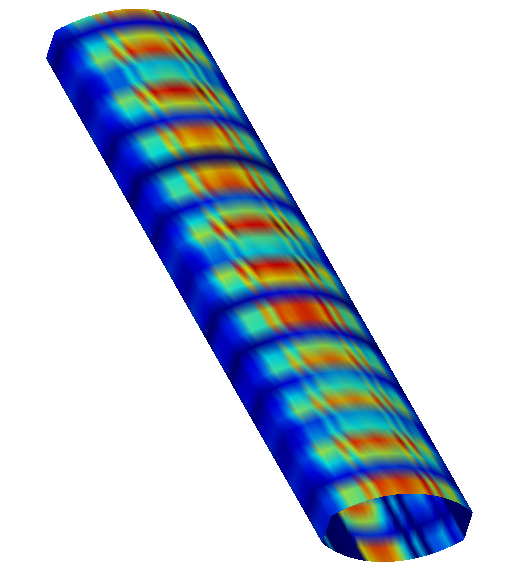}
\caption{Snapshot of a Vorpal simulation showing propagation of a TE wave through the CesrTA
beam pipe}
\label{fig:vorpalpic}
\end{center}
\end{figure}

\section{Electron Cloud induced Phase shift from Transmission Through 
Field-Free regions and the effect of reflections} 
The first experiments using microwaves to assess the cloud densities
involved simply transmitting the wave using a beam position monitor (BPM)
and receiving the transmitted wave from another BPM downstream to the
traveling wave \cite{Caspers,StefanoPRL}. As shown in Ref~\cite{SonnadPAC}, 
the electron cloud induced phase shift per unit length of transmission
in the absence of external magnetic fields and a uniform electron
distribution, can be related to the electron density as follows,
\begin{equation}
 \Delta \phi = \frac{\omega_p^2}{2c(\omega^2 - \omega_{co}^2)^{1/2}}
\label{phaseshift}
\end{equation} 
where $\omega_{co}$ is the angular cutoff frequency for a waveguide in vacuum, 
$\omega_p = \sqrt{n_e e^2/\epsilon_0m_e}$ is the angular plasma frequency,
with $n_e$ the electron number density, $e$ the charge of the electron,
$c$ the speed of light, $m_e$ the electron mass and $\epsilon_0$ the
free space permitivity. In Ref~\cite{SonnadPAC}, this formula was validated 
through simulations for a square cross section beam pipe.  The derivation of the phase shift 
given by Eq~\ref{phaseshift} used the dispersion relationship given in Ref~\cite{Uhm}, 
\begin{equation}
  k^2 = \frac{\omega^2}{c^2} - \frac{\omega_p^2}{c^2} - \frac{\omega_{co}^2}{c^2}
\label{disprel} 
\end{equation}
which was proved to be valid for a circular waveguides free of external
magnetic fields. The results shown in this paper validates the formula for 
a Cesr beam pipe geometry. All these results collectively indicate that the 
formula is valid for any type of geometry. In the Appedix, we provide an explicit 
proof that this is indeed the case. The Cesr beam pipe geometry may be represented 
in the form of two circular arcs (radius 0.075m) connected with flat side planes.
It is about 0.090m from side to side and 0.050m between the apices of the arcs. 
The cutoff frequency of lowest mode for this geometry is known to be 
around $1.888$ GHz from past experiments and calculations related to the beam pipe.  
Figure~\ref{fig:vorpalpic} provides a snapshot of the simulation, showing
the propagation of a wave through the Cesr beam pipe. 
\begin{figure}[ht]
\begin{center}
\includegraphics*[width=0.8\columnwidth]{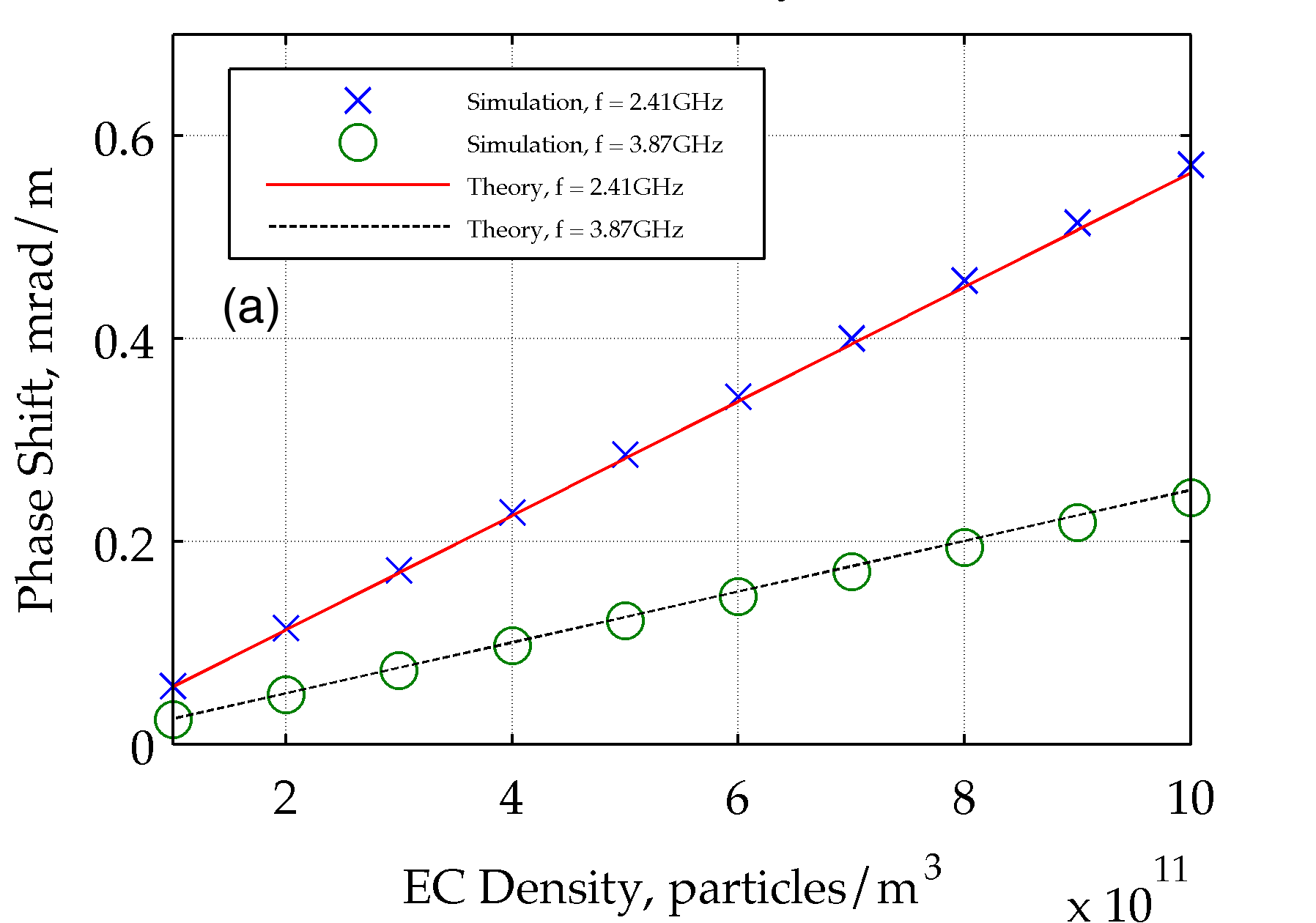}
\includegraphics*[width=0.8\columnwidth]{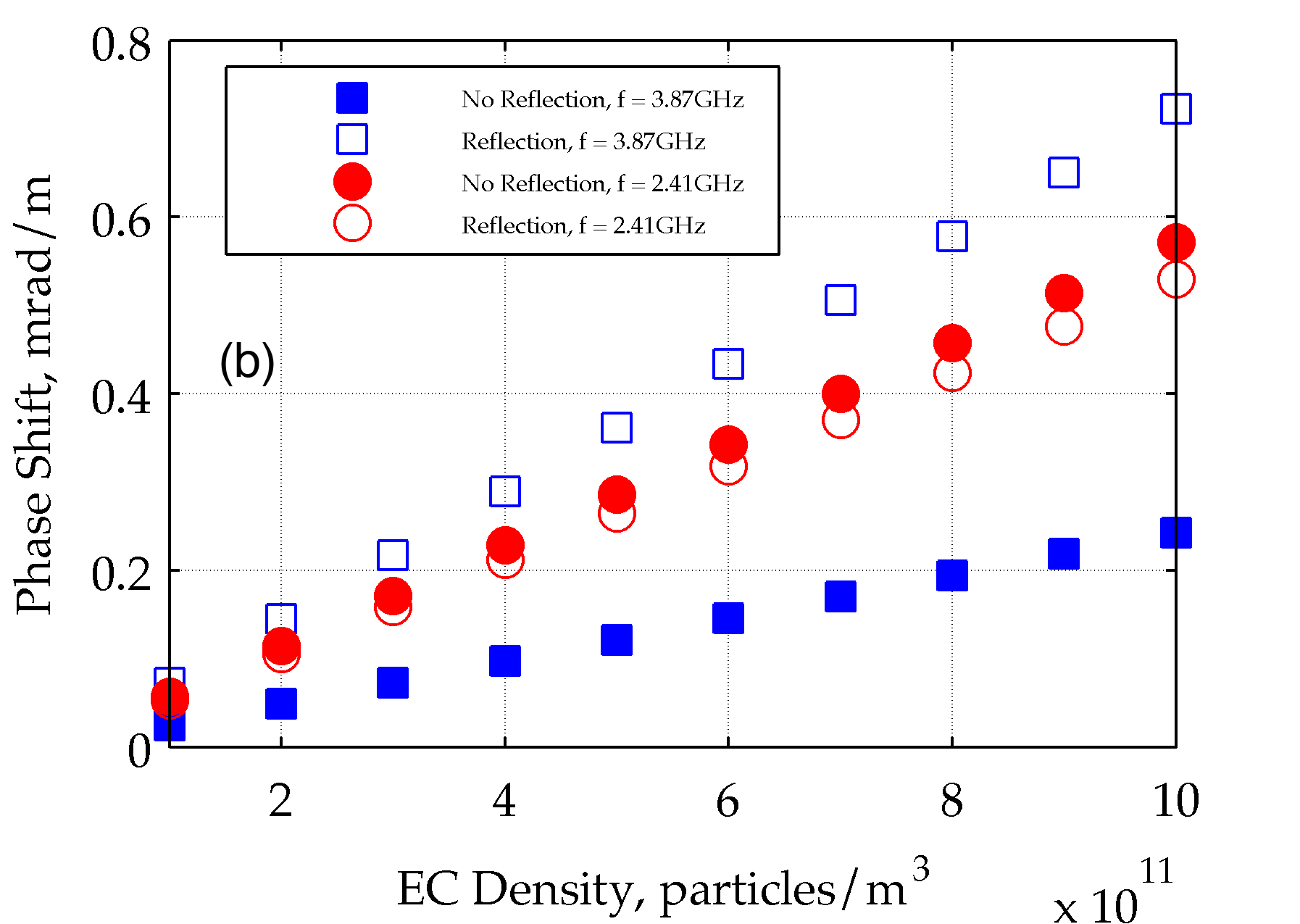}
\caption{Variation of (a) phase shift with cloud density for a Cesr beam pipe geometry,
and (b)the effect of reflections on the phase shift}
\label{fig:phaseshift1}
\end{center}
\end{figure}

Calculation of electron induced phase shift was performed through separate simulations 
of the wave transmission through a vacuum beam pipe and through a beam pipe with electrons 
respectively. At a certain axial distance $L$ from the location at which the wave was 
launched, the variation of the voltage between the midpoints of the top and bottom boundaries 
of the beam pipe cross section were recorded as a function of time. After normalizing the 
amplitudes of the two waves to unity,  their difference gives a sinusoidal wave with an 
amplitude equal to the phase shift between the waves. Suppose that the angular frequency of 
the wave is $\omega$ and phase shift is $\delta$. The phase shift for nominal cloud densities is 
small enough that $\sin(\delta) \approx \delta$. Hence we have $\cos(\omega t) - \cos(\omega t + \delta) 
\approx \delta \sin(\omega t + \delta/2)$. The amplitudes of all the
waves were calculated from their respective numerical RMS values. 

To confirm the relationship between the phase shift and electron cloud
density, simulations were done with a Cesr beam pipe with a length of 0.5m.
Figure~\ref{fig:phaseshift1}(a) shows that simulations agree
well with the analytically predicted values given by
Eq~\ref{phaseshift}. This establishes the accuracy of the simulation method
as well as the validity of Eq~\ref{phaseshift} for any geometry. We see 
that the electron cloud induced phase shift increases as one
approaches the cutoff frequency. While this is desirable because it
amplifies the modulation side-bands relative to the carrier signal, one will
encounter reduced transmission as the carrier frequency approaches the
cutoff. 
\begin{figure}[ht]
\begin{center}
\includegraphics*[width=0.9\columnwidth]{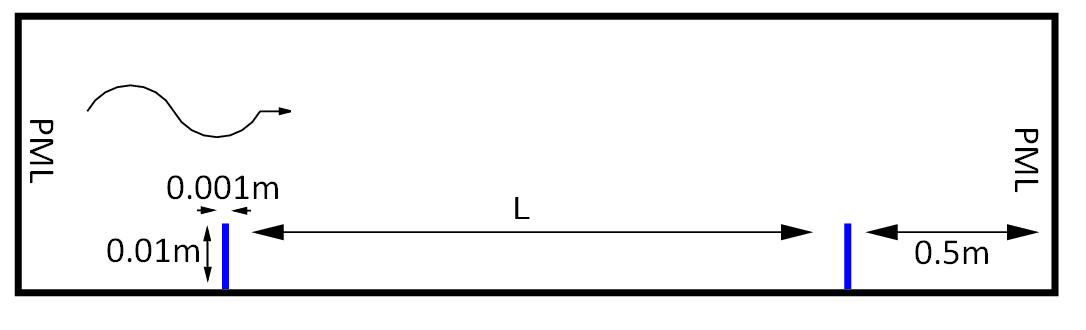}
\includegraphics*[width=0.9\columnwidth]{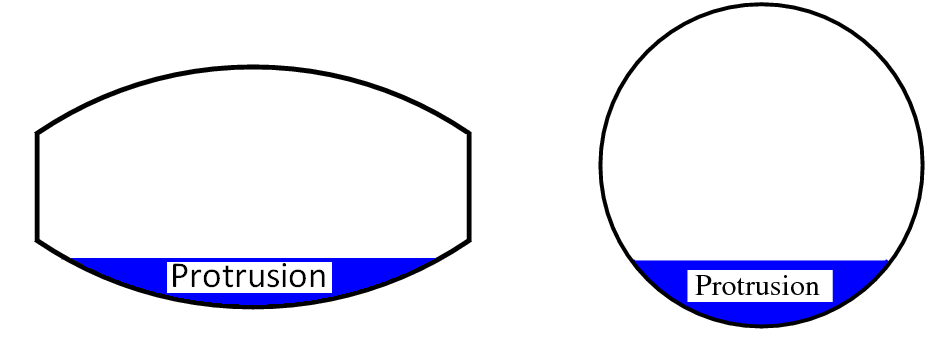}
\caption{A schematic of the simulations with protrusions serving as partial
reflectors, showing the Cesr and a circular cross section beam pipes. The
latter geometry is used in Sec. 3}
\label{fig:protrusions}
\end{center}
\end{figure}

Due to the presence of several mechanical and electronic components 
all along the vacuum chamber, it is not very likely that one can perform
phase shift experiments that are entirely free of internal reflections.
As discussed in Ref~\cite{DeSantisPRSTAB}, in an experiment, these internal 
reflections would potentially affect the value of the phase shift.  
A wave reflected from a device that lies beyond the segment being measured, 
would be received by the detector when it comes back after reflection. 
At the same time, waves could get reflected back and forth within
the segment before eventually being received at the detector. In both
cases, the reflected wave would have sampled a length different from
that meant to be sampled and would thereby contaminate the signal,
because the phase shift is proportional to the length of transmission. In
order to understand this effect, the simulations were altered to include
two protruding conductors, which would reflect some of the transmitted
wave. The protrusions were slabs in the transverse plane, extending from
the bottom to 1cm above the apex of the lower arc (see
Fig.~\ref{fig:protrusions}). They were spaced 0.4 meters apart, including 
the thickness of the protrusions, which was 1mm.  The frequencies used for 
this study 2.41 GHz and 3.87 GHz, the same as those shown in
Fig~\ref{fig:phaseshift1}(a), correspond to  the resonant harmonics ($n =
4$ and $n = 9$, respectively) of a 0.4 meter ``resonant cavity". This
was done in order to maximize reflections. Fig.~\ref{fig:phaseshift1}(b)
shows the resulting phase shifts from these calculations. The solid shapes
represent the data for no reflections and are the same data that appear
in Fig.~\ref{fig:phaseshift1}(a).  The open shapes represent the phase
shifts in the presence of reflection. These results clearly indicate
that internal reflections modify the expected phase shift. The nature
of the alteration of phase shift depends upon the complexities of the
transmission-reflection combination, and the instrumentation used for
the method. However, the results show that the linear relationship between 
phase shift and electron density is always preserved.

\section{Standing Waves from Partial Reflectors and Electron Cloud Induced 
Resonant Frequency Shift}
While internal reflections may interfere with phase shift measurements,
they can also be used to trap a wave. This trapped wave can be used 
to measure the electron cloud density, as discussed in Ref~\cite{SikoraIPAC11}. 
This section shows the results of numerical simulation of such an
experiment. The geometries used were (a) Cesr beam pipe also used in the 
previous section and (b) beam pipe with a circular cross section of radius 4.45cm. 
Both of them had conducting protrusions that were 1mm thick and 1cm high from the 
base as shown in Fig~\ref{fig:protrusions}. The cutoff frequency for the
circular beam pipe can be calculated from the analytic expression, and is 1.9755GHz
for the lowest ($TE_{11}$) mode. 
\begin{figure}[htb]
  \centering
  \includegraphics*[width=0.9\columnwidth]{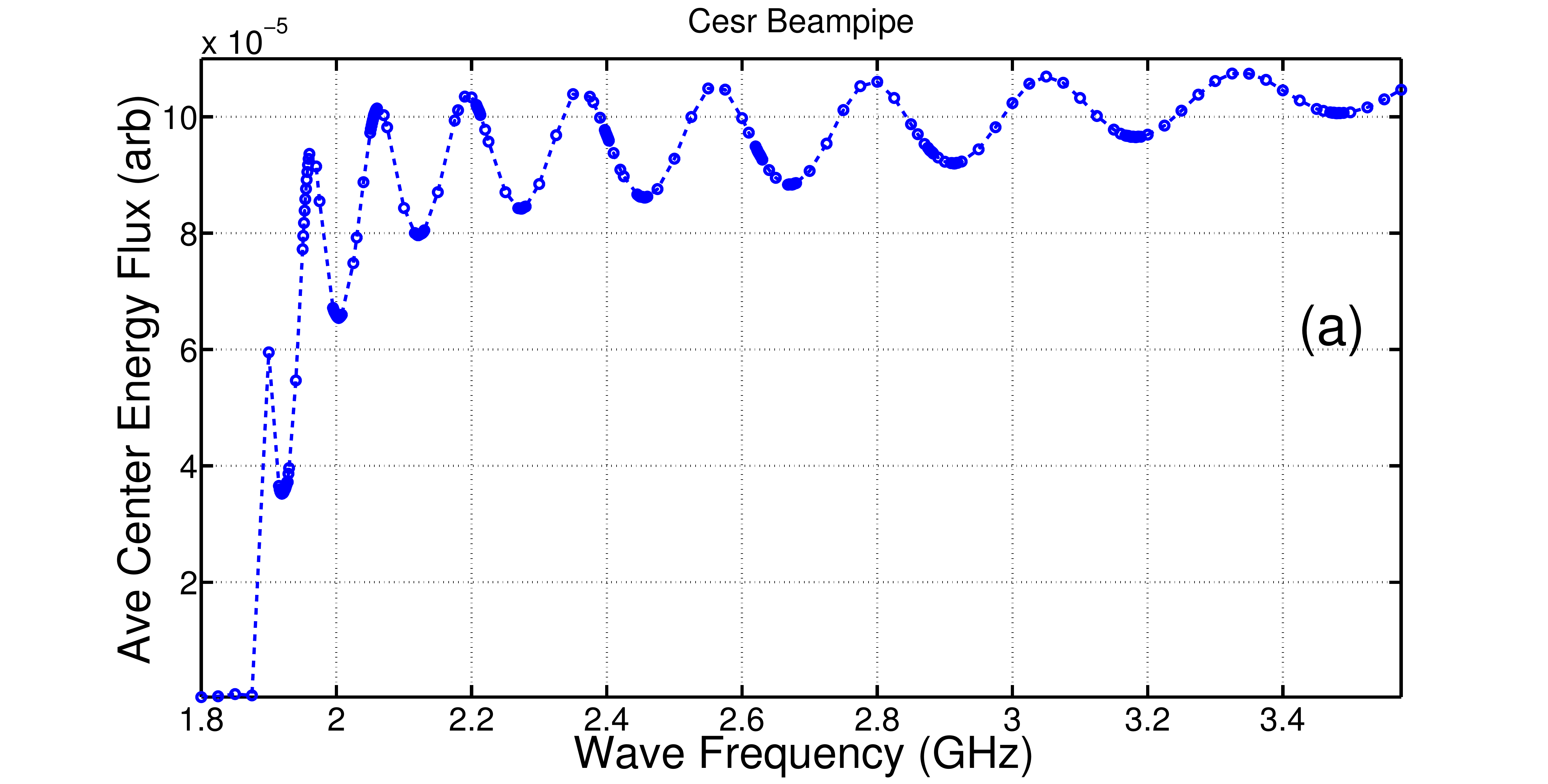}
  \includegraphics*[width=0.9\columnwidth]{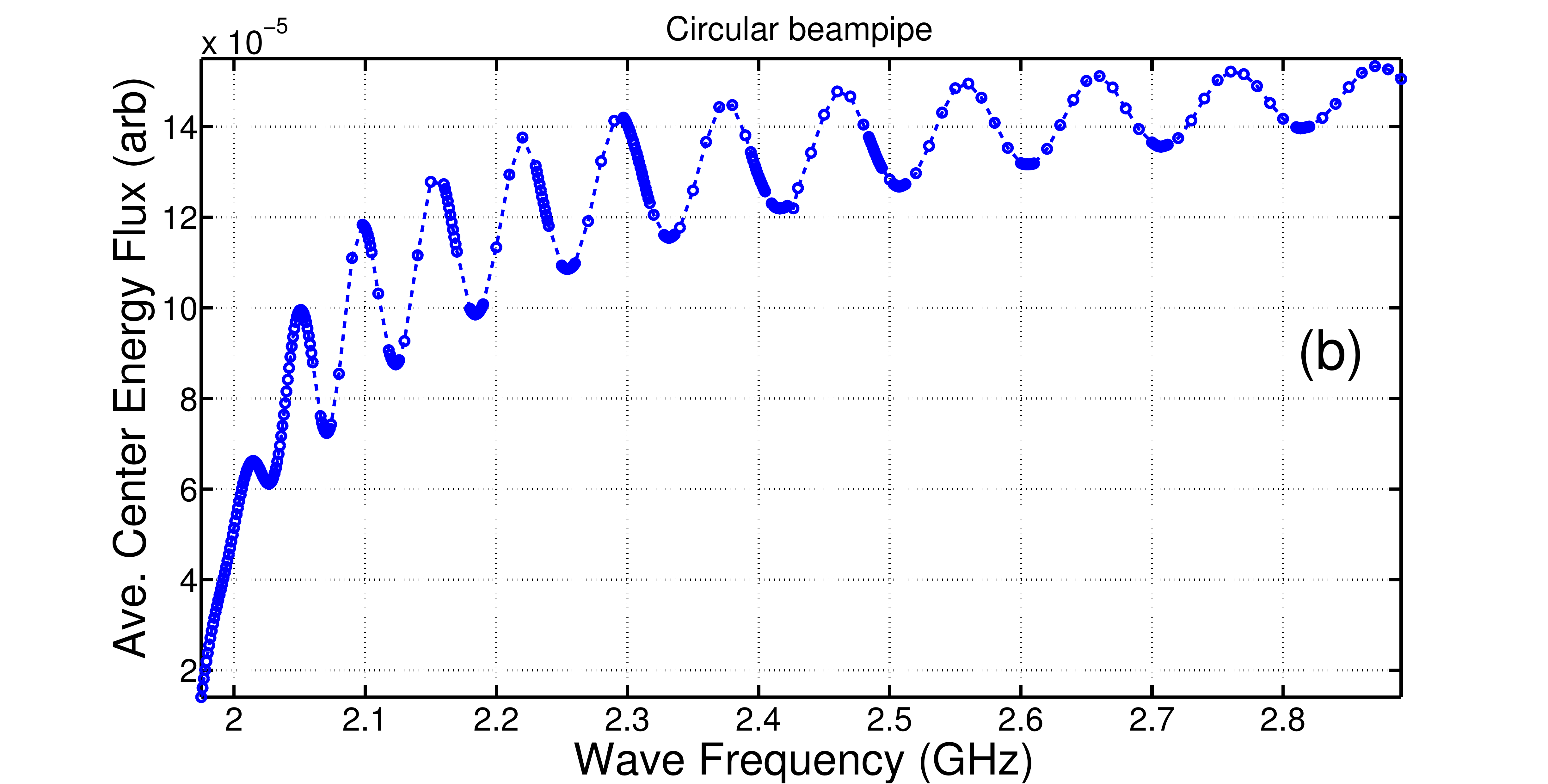}
  \caption{The average energy flux calculated over a range of frequencies, with the
local minima indicating resonance points for (a) the Cesr beam pipe, (b) beam pipe
with circular cross section}
\label{fig:poyntingflux}
\end{figure}

A trapped mode results when standing waves are induced between the reflectors. 
In order to test the
effectiveness of inducing such a standing wave using partial reflectors,
simulations were done with an empty wave guide over a large range of
frequencies. It should be noted that, since there is only a partial
reflection of the wave taking place, there is always a net transmission
of energy across the segment between the protrusions.  The wave energy
that escapes the partial reflectors gets absorbed into the PML regions.
Identification of a resonance was done as follows. At each frequency and 
each time step, the wave energy flux was computed by integrating the 
Poynting vector across a plane located at the mid point between the two 
protrusions. This plane was oriented transverse to the axis and covered the 
entire cross section. For each of these frequencies, the mean of the energy flux 
was calculated over the period of the simulation. It is expected that as the 
frequency approaches that of a standing wave, this averaged flux would reach a 
local minimum. This is because of increased "back and forth" transmission which 
does not contribute to the average flux due to cancellation.  

Figure~\ref{fig:poyntingflux} shows the average energy flux calculated for a 
variety of frequencies spanning over several resonance points for (a) the Cesr
beam pipe (b) the circular cross section beam pipe. The local minima seen on these 
plots correspond to a standing wave mode. The length of the section, including 
the width of the protrusions was 0.4m for the Cesr beam pipe. The length of the
circular beam pipe, including the width of the protrusions was 0.88m 

\begin{figure}[htb]
  \centering
  \includegraphics*[width=0.9\columnwidth]{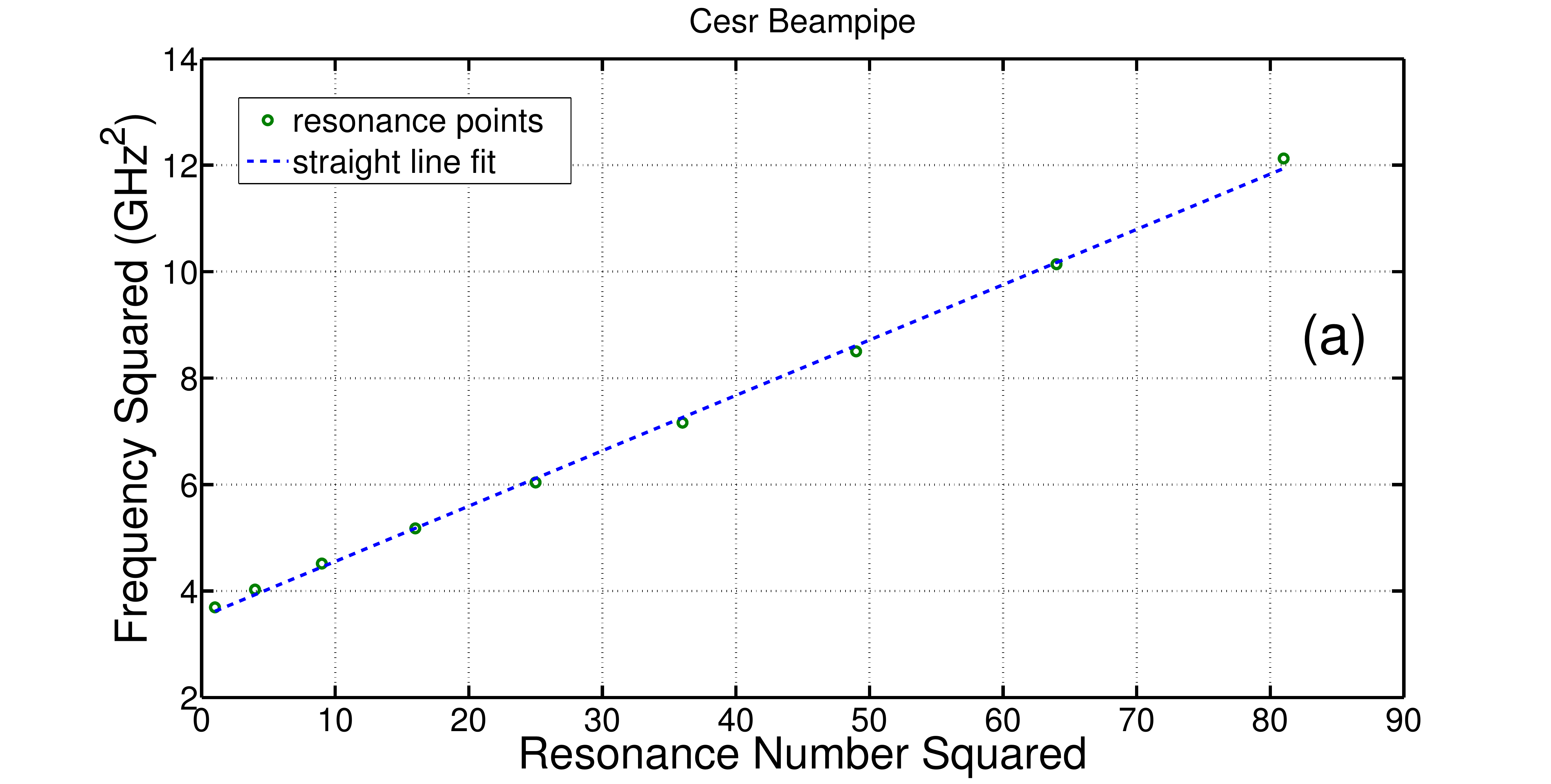}
  \includegraphics*[width=0.9\columnwidth]{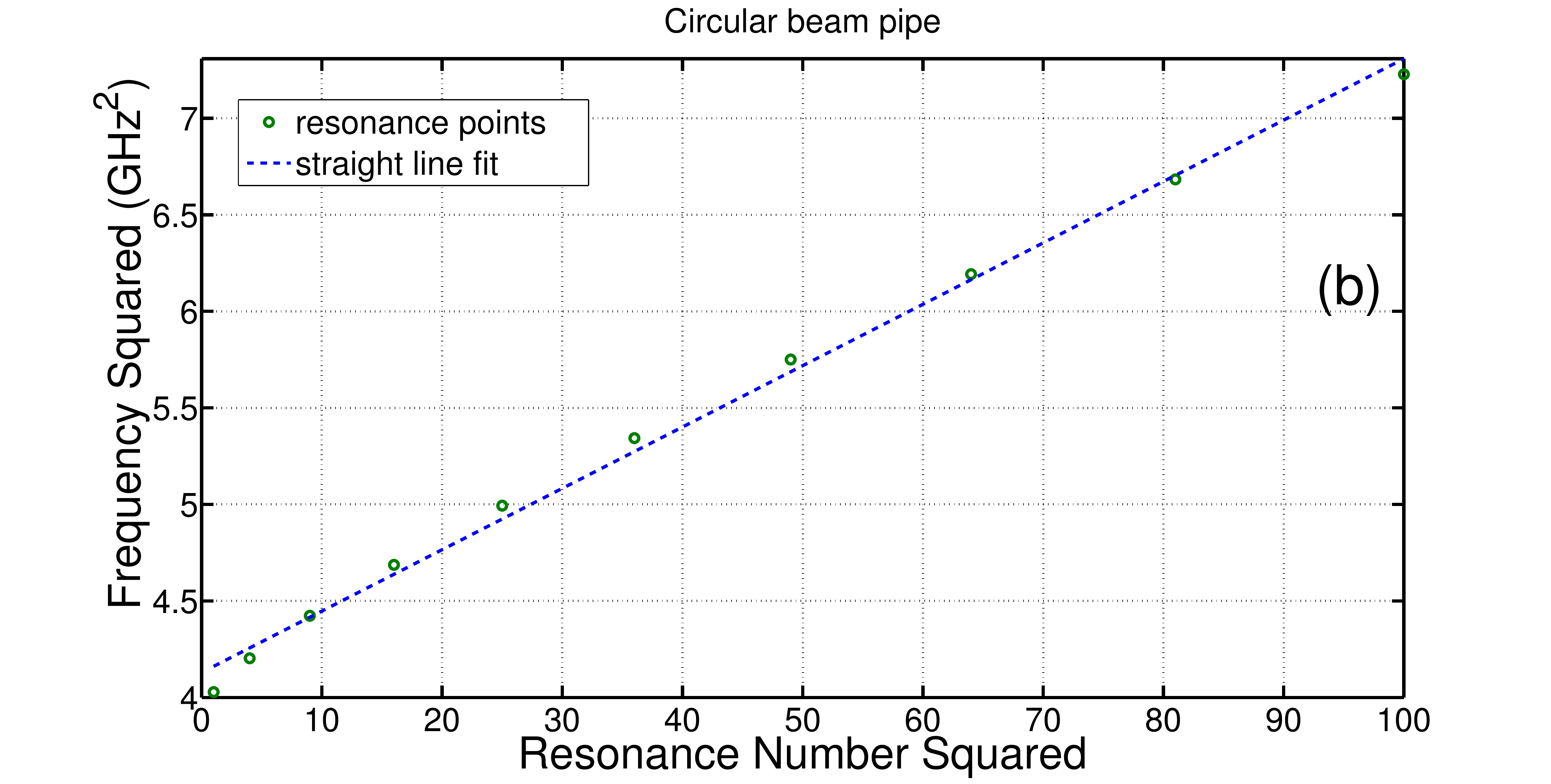}
  \caption{Resonance points obtained from the data in Fig~\ref{fig:poyntingflux}
showing the linear relationship between $f_0^2$ and $n^2$ according to Eq~\ref{vacstandwave}}
\label{fig:f2vsn2}
\end{figure}

A standing wave occurs when the wavelength $\lambda$ and the length
$L$ of the segment are related such that for any integer $n$, $L = n\lambda/2$. 
The dipersion relationship of a waveguide with wave frequency $f_0$ and cutoff 
frequency $f_{co}$, is given by $c^2/\lambda^2 = f_0^2 - f_{co}^2$. Expressing $\lambda$ 
in terms of $L$, for the standing wave, this then gives
\begin{equation}
 f_0^2 = \frac{c^2}{4L^2}n^2 + f_{co}^2
\label{vacstandwave}
\end{equation}   
This indicates the linear relationship between  $f_0^2$ and $n^2$, and also
the relationship of $L$ with the slope, and $f_{co}$ with the intercept of the
straight line. Equation~(\ref{vacstandwave}) was used to confirm that the local 
minima in Fig~\ref{fig:poyntingflux} correspond to resonance points. 
Figure~\ref{fig:f2vsn2} shows the value of $f_0^2$ plotted as a function of the 
corresponding value of $n^2$ for (a) the Cesr beam pipe and (b) beam pipe with
a circular cross section. Performing a straight line fit on these points
for case (a), yielded the relationship $f^2(GHz^2) = 0.104n^2 + 3.515$. This gives 
$L = 0.465m$ and $f_{co} = 1.875GHz$. For case (b), a similar operation gives
$f_0^2(GHz^2) = 0.0318n^2 + 4.1276$ giving $L = 0.841m$ and $f_{co} = 2.0316GHz$.  
These values are close enough to the expected ones and ascertain the accuracy 
of such a method in determining standing waves between partial reflectors. 

The presence of an electron cloud would result in a shift in the 
standing wave frequency. Experimentally, it is possible to measure 
this in the form of frequency modulation side-bands associated with 
the periodic passage of a train of bunches creating electron clouds.
Using, Eq~(\ref{disprel}) we can show that the condition for standing 
waves given by Eq~(\ref{vacstandwave}) is modified by an electron cloud
as follows,
\begin{equation}
 f_e^2 = \frac{c^2}{4L^2}n^2 + f_{co}^2 + f_p^2
\label{elecstandwave}
\end{equation}
where the wave frequency is now denoted by $f_e$ and $f_p$ is the plasma 
frequency. Subtracting Eq~\ref{vacstandwave} from Eq~\ref{elecstandwave}, and
in the limit of small frequency shifts, we get, 
$f_e^2 - f_0^2 \approx 2\Delta ff_0 = f_p^2$, where $\Delta f = f_e - f_0$. 
On inserting the expression for $f_p$, this then gives a simple 
expression relating the shift in resonant frequency as a function of electron density.
\begin{equation}
  \Delta f = \frac{n_{e}e^2}{8\epsilon_0 m_e\pi^2f_0}=\frac{n_e r_e c^2}{2\pi f_0}
\label{freqshift}
\end{equation}
in which $r_e$ is the classical electron radius. This shows that the frequency 
shift is proportional to the electron cloud density. 

An effort is underway at CesrTA to use this method to measure the density of the 
electron cloud within the beam pipe section where the reflections are occurring 
\cite{SikoraNIM}. Thus, it became necessary to test this phenomenon with simulations. 
Simulation of the frequency shift of standing waves was done for a Cesr beam pipe
cross section as well as a beam pipe with a circular cross section. All the parameters 
were the same as before except that the length of the section between the partial 
reflectors for the Cesr beam pipe was modified from 0.4m to 0.9 m, which was somewhat 
close to one the setups under study at CesrTA. The length of the circular cross section was 
retained at 0.88m. Since the frequency shift induced by electrons is very small, it 
is required that the resonant frequency be determined accurately. To do this, a parabolic 
fit was made to the averaged energy flux in the vicinity of the minimum point, using the
available points obtained from simulation. The expression for the parabola may 
be obtained from a Taylor expansion of the function around the minimum. This gives
the mean energy flux as a function of frequency $f$ near the $n$th minimum point $f_n$.
Thus, we have 
\begin{equation}
 {\cal E}(f) = {\cal E}^{\prime \prime}(f_n)f^2 - {\cal E}^{\prime \prime}(f_n)f_nf + 
[{\cal E}(f_n) + {\cal E}^{\prime \prime}(f_n)f_n^2]
\end{equation}
where ${\cal E}(f)$ is the averaged energy flux, ${\cal E}^{\prime \prime}(f_n)$ is the second
derivative of ${\cal E}$ evaluated at $f_n$. The first derivative
of ${\cal E}$ at the minimum point vanishes. Using the computed coefficients of the parabola
one can solve for $f_n$. 

\begin{figure}[htb]
   \centering
  \includegraphics*[width=0.9\columnwidth]{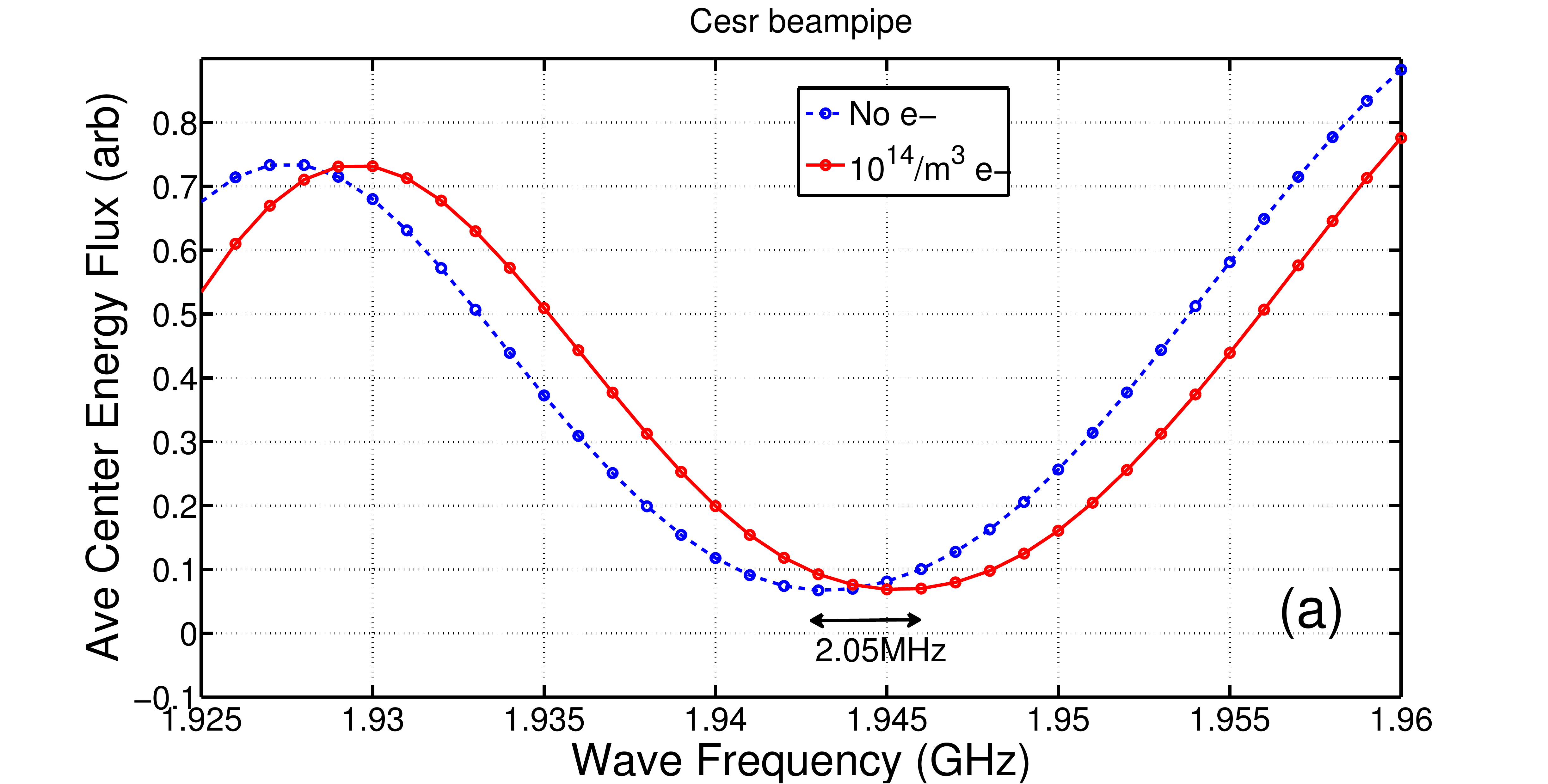}
  \includegraphics*[width=0.9\columnwidth]{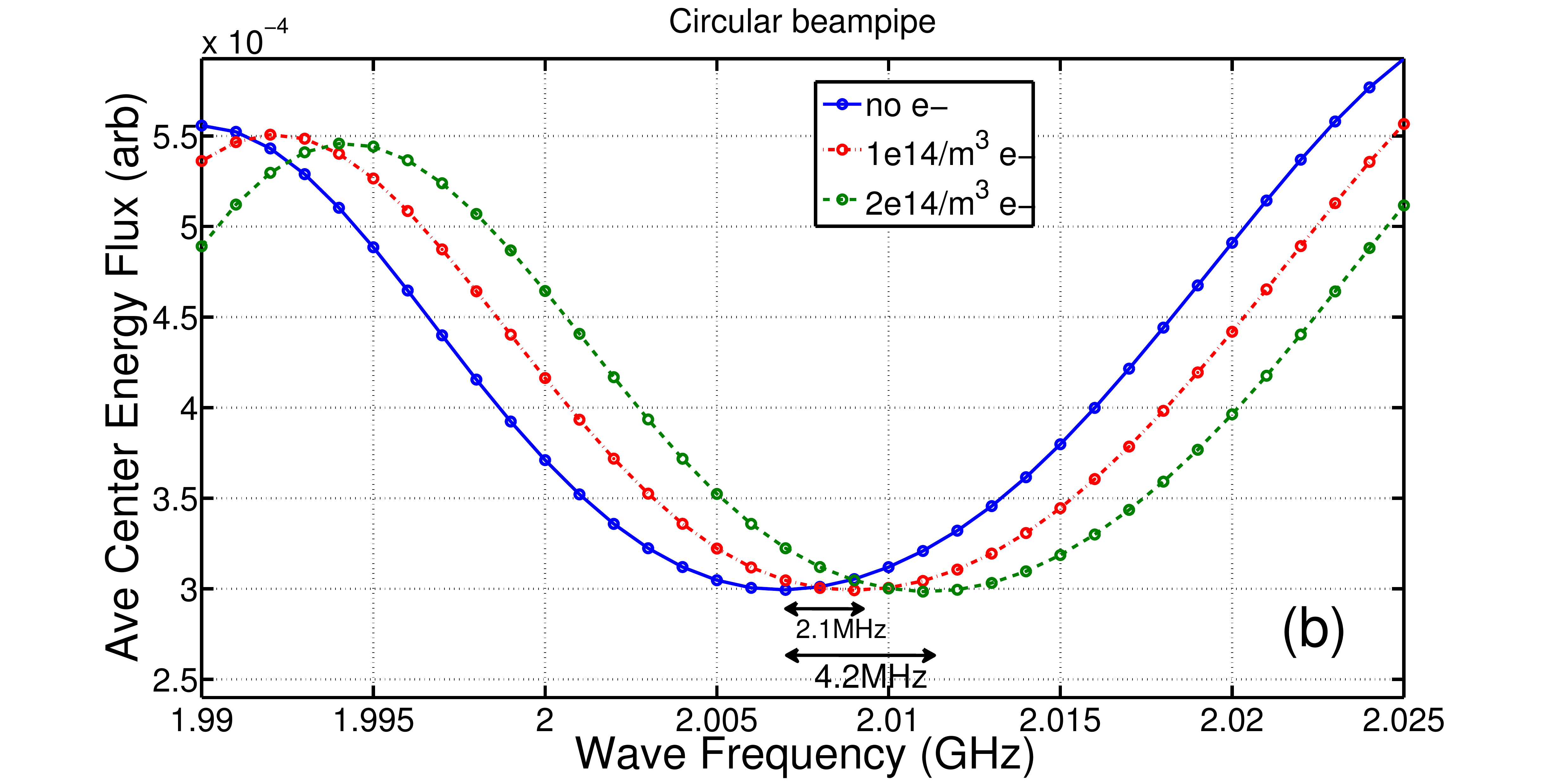}
   \caption{Simulation results showing a shift in the $n=2$ resonance frequency
induced by electron clouds in (a) the Cesr beam pipe and (b) the circular
cross section beam pipe.}  
\label{resshift}
\end{figure}

Figure~\ref{resshift} shows the frequency shift induced by electron
clouds for the $n=2$ mode for (a) the Cesr beam pipe and the (b) the circular
beam pipe. For case (a), the electron density used in this calculation was 
$10^{14}$ m$^{-3}$, which is rather high, but helps validate Eq~(\ref{freqshift})
with simulations more accurately. For these parameters, the $n=2$
resonance occurs at 2.0033 GHz and the expected frequency shift due to
electrons is 2 MHz. Simulations show a shift of 2.05 MHz. For case (b),
we used two values of electron densities, $10^{14}$ m$^{-3}$ and
$2 \times 10^{14}$ m$^{-3}$. The expected frequency shift for the 
$10^{14}$ m$^{-3}$ case is 2.013MHz. The simulated frequency shift for
this was 2.1MHz, and for an electron density of $2 \times 10^{14}$ m$^{-3}$, it 
was 4.2MHz. Thus we were able to establish that reasonably accurate values of 
frequency shift for such an experiment may be determined from simulations. 
It is interesting to note that the error obtained in the resonant frequency 
itself was around $2MHz$, however the shift induced by electron clouds 
always had reasonable agreement with Eq~\ref{freqshift}. The agreement with 
theory provides confidence in estimating electron densities based on Eq~\ref{freqshift}
from  measurements. Typical electron cloud densities are of the order
of $10^{12}$m$^{-3}$ leading to  frequency shifts about a factor of 100 
smaller than those simulated here. Simulating frequency shifts this small
would in principle be possible by scaling all numerical parameters appropriately, 
but this would have been a far more intensive computational process. However, it 
is well known that in practice, such small frequency modulations can be easily 
measured with standard spectrum analyzers. 

\section{Transmission and Phase Shift Through Dipole and Wiggler Field Regions}
This section discusses the effect of external dipole and wiggler magnetic fields
on the phase shift measurements. Simulations done in the past, have revealed 
that the phase shift is greatly amplified in the presence of an
external magnetic field if the electron cyclotron frequency lies in the vicinity
of the carrier frequency \cite{SonnadAPS, VeitzerDOE}. Following these results, the 
cyclotron resonance was soon confirmed at an experiment performed at the SLAC 
chicane \cite{PiviEPAC}. The SLAC chicane has since been transferred to CesrTA, 
where these studies continue to be made \cite{DeSantisPAC11}. 
 
In the presence of an external magnetic field and electron clouds, the
medium is no longer isotropic and the polarization of the transmitted
microwave plays an important role in the outcome of the measurement.
When the wave electric field is oriented perpendicular to the external
magnetic field, the mode is referred to as an Extraordinary wave or
simply X-wave. In this situation, if the external dipole field corresponds
to an electron cyclotron frequency close to the carrier wave frequency,
we see an enhanced phase shift. The phenomenon is well understood in the
case of open boundaries. It is usually referred to as upper hybrid resonance.
The dispersion relationship for the open boundary case is given as
follows (see for example ref~\cite{GoldstoneRutherford}).
\begin{equation}
\frac{c^2k^2}{\omega^2} = 1 - \frac{\omega_p^2(\omega^2 - \omega_p^2)}
 {\omega^2(\omega^2 - \omega_h^2)}
\label{openbdryres}
\end{equation}
The quantity $\omega_h$ is the upper hybrid
frequency which is given by $\omega_h^2 = \omega_p^2 + \omega_c^2$, $\omega_c = eB/m_e$ being the
electron cyclotron frequency for the given magnetic field $B$. When $\omega_h \to \omega$,
it is clear that $k \to \infty$. It can also be seen that as $\omega_h \to \infty$, i.e., for
very high magnetic fields, the relationship between $k$ and $\omega$ approaches that of
propagation through vacuum. In the case of electron clouds in beam pipes, the plasma frequency is of
the order of a few 10 MHz while the carrier frequency is around 2 GHz. In this regime, 
it is reasonable to state that resonance occurs when $\omega_c \to \omega$. Since the phase advance is the 
product of the wave vector $k$ and the length of propagation, we see that the electron cloud induced 
phase shift will theoretically go to infinity.  Equation~(\ref{openbdryres}) is not valid for waveguides, 
which have finite boundaries. Nevertheless, simulations show that the same qualitative features are 
exhibited also for propagation through waveguides. 

\begin{figure}[ht]
   \centering
   \includegraphics*[width=70mm,height=40mm]{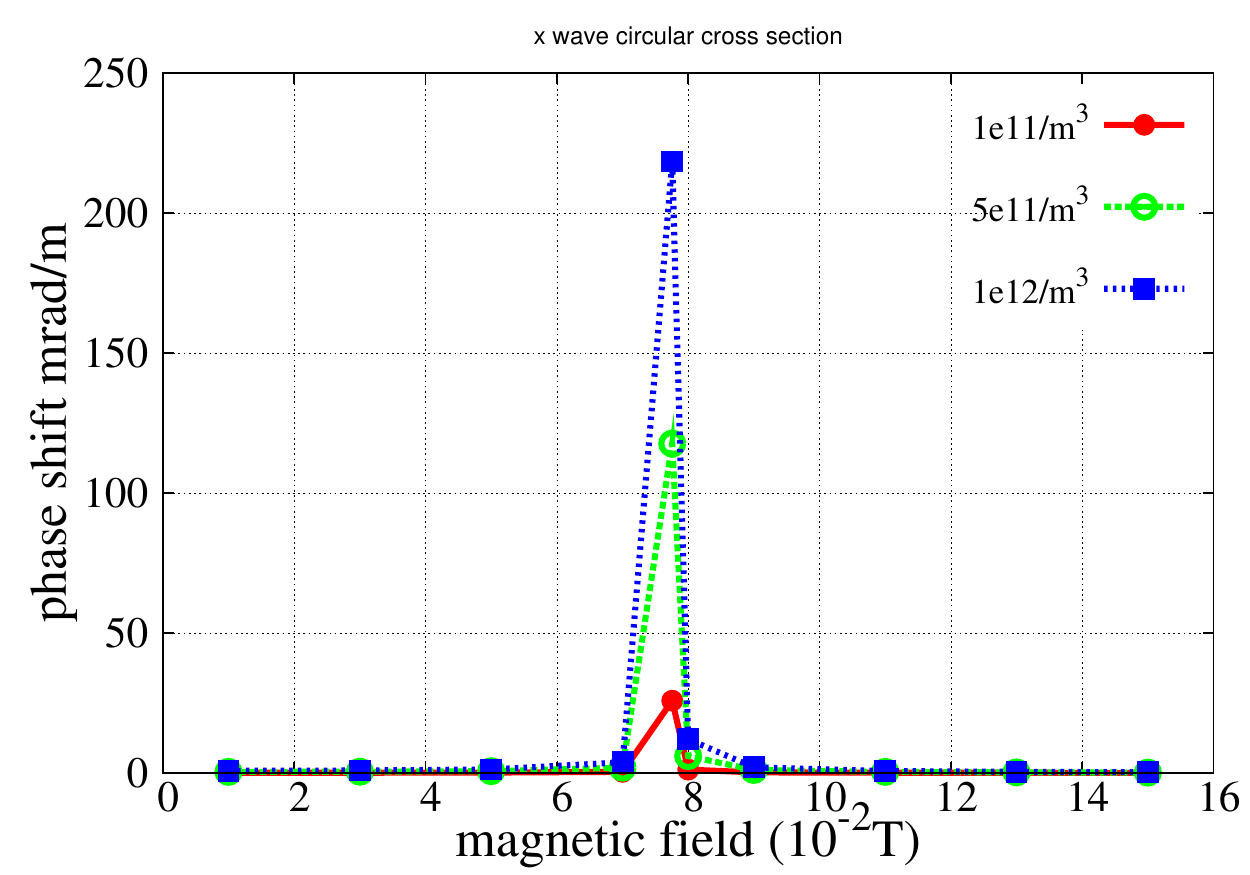}
   \includegraphics*[width=70mm,height=40mm]{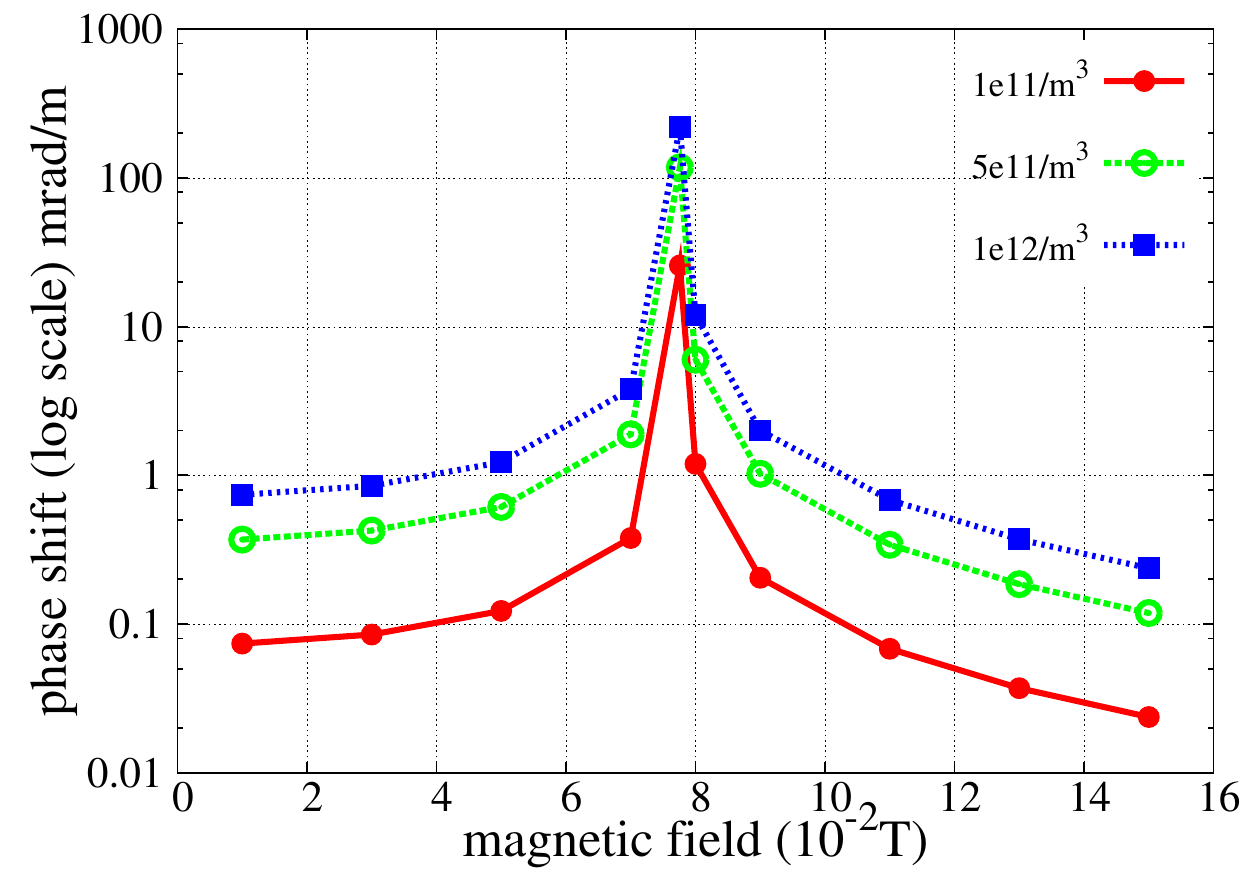}
   \caption{Phase shift vs. magnetic field for dominant X wave
propagation for different cloud densities. The peak corresponds to a cyclotron resonance.}
   \label{xwave}
\end{figure}

Figure~[\ref{xwave}] shows the enhanced phase shift for three values of cloud densities when the 
cyclotron frequency approaches the carrier frequency. The beam pipe cross section was circular with 
a radius of 4.45 cm, which leads to a cutoff at 1.9755GHz at the fundamental $TE_{11}$ mode. These 
parameters match with the beam pipe geometry of the PEP II/CesrTA chicane section. The wave frequency 
used in the simulation was 2.17 GHz. The magnetic field corresponding to this cyclotron frequency is 
0.077576T. The wave was excited with the help of a sinusoidally varying electric field pointing perpendicular 
to the external magnetic field. 

\begin{figure}[htb]
   \centering
   \includegraphics*[width=70mm,height=40mm]{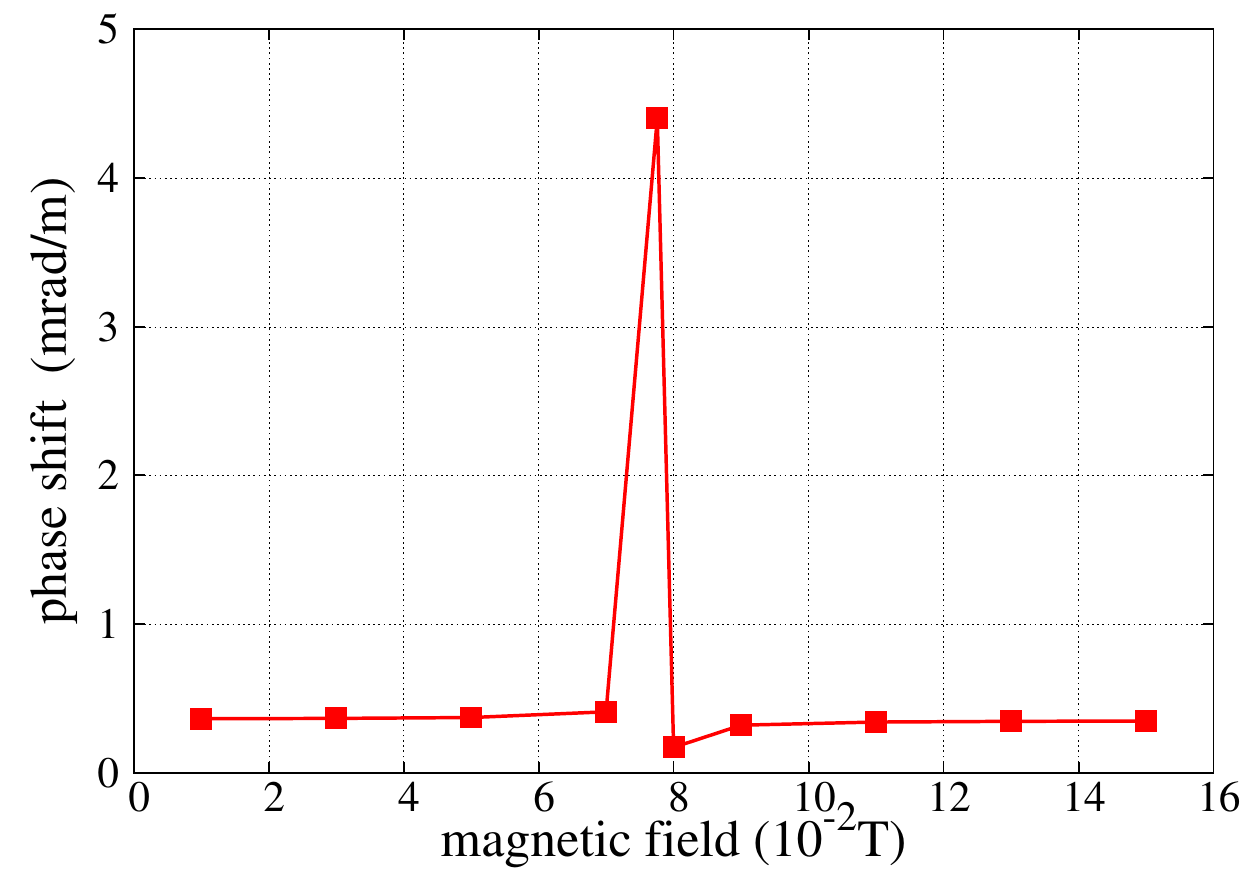}
   \caption{Wave launched with a dominant O wave component showing weak resonance}
   \label{owave}
\end{figure}

When the  wave is polarized
with an electric field that is parallel to the external magnetic field, the mode is
often referred to as the Ordinary wave, or the O-wave. In this case one would not expect 
any effect created by the external magnetic field. This would be the case for a rectangular 
cross-section waveguide, or with open boundaries. However, with a circular cross section as in
our study, the boundary conditions would force a component perpendicular to the magnetic
field in the  wave electric field even if it is launched with a purely parallel electric field.
Thus, it is inevitable that a weak component of a wave with an orthogonal polarization gets excited.
Additionally, the method employed in launching the wave in the simulations is similar to that 
performed in experiments, where an electric field wave is excited along a particular direction 
over a surface area. The functions describing the wave for
a cylindrical geometry are Bessel functions involving the radial and azimuthal variables. Unless care 
is taken to excite a wave having the given functional form, the wave is expected to couple 
itself to two orthogonal modes with varying degrees of intensity. Additionally, the detection 
system, would receive the effect of the two modes to varying degrees. Disentangling 
this combination will involve more analysis, guided by simulations. Due to these effects,
we see a weak resonance effect even in a wave excited with a purely vertical electric field,
that is aligned to the external magnetic field as indicated.  Figure~\ref{owave} shows the
presence of such a weak resonance and this effect has been observed at CesrTA as well.
\begin{figure}[htb]
  \centering
  \includegraphics*[width=70mm,height=40mm]{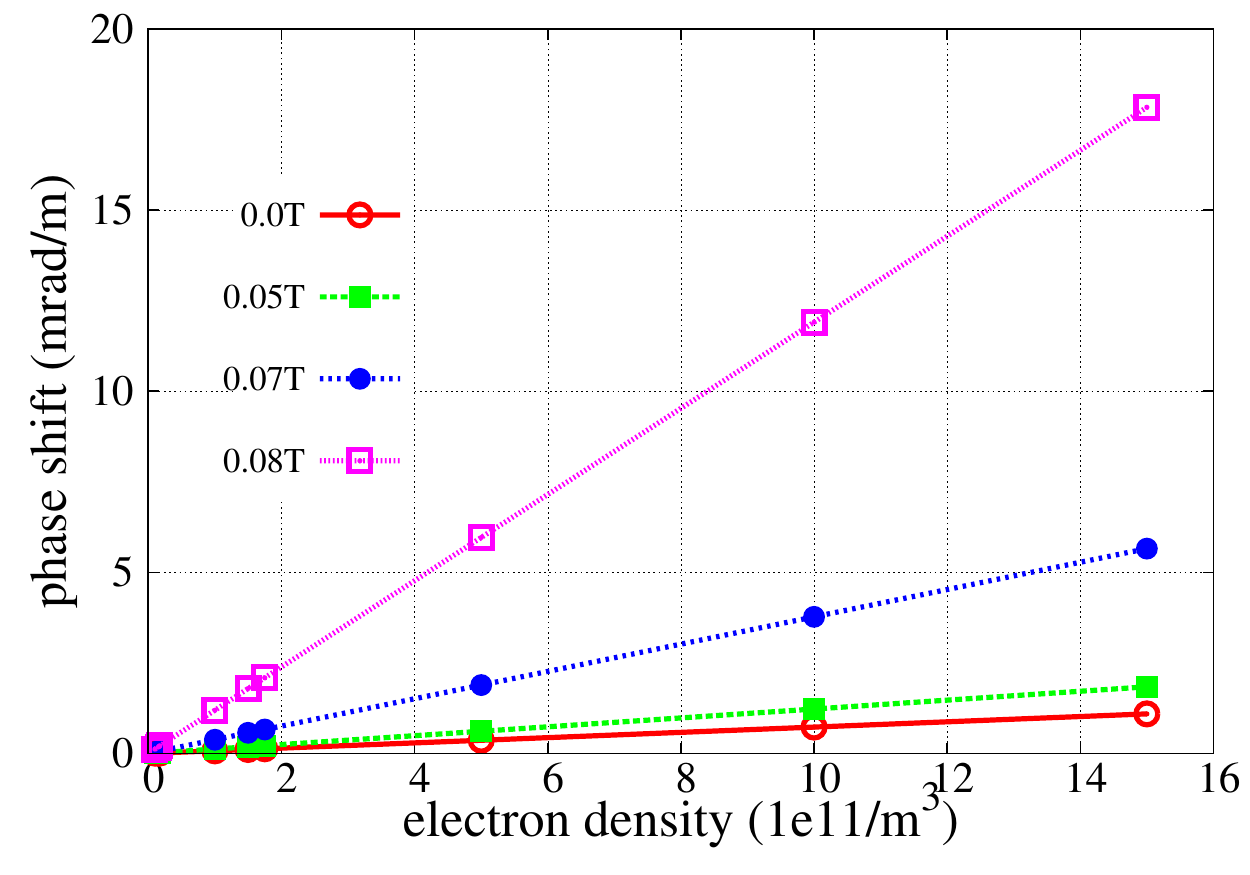}
  \caption{Variation of phase shift with cloud density for different
  magnetic fields for dominant X-wave propagation.}
  \label{dens_scan}
\end{figure}

Figure~[\ref{dens_scan}] shows the variation of phase shift with electron cloud density
at different settings of external magnetic fields. In these simulations, the wave
was excited with an electric field perpendicular to the external magnetic field. 
These densities are typical of what is produced in CesrTA. The plots show that the 
variation of phase shift with density remains linear even when one is close to 
resonance. This is expected to be true as long as the plasma frequency is much smaller than 
the wave frequency, regardless of how complex the dispersion relationship of the wave is. Thus, 
one could easily amplify the signal with the help of an external magnetic field to 
monitor relative changes in cloud density, if not the absolute density. 

Experiments have been done to study the phase shift across the damping
wigglers at CesrTA. These experiments correspond to various bunch currents
and wiggler field settings. The wiggler field setting influences the measurement
in more than one way. The wiggler field affects the motion of the electrons, which 
influences the secondary production of the cloud. The synchrotron radiation
flux is determined by the strength of the wiggler field, and this in turn determines
the photoemission rate of the cloud. Both these effects determine the density of 
the cloud. The electron density is not uniform across the length of the wiggler, as 
shown in Ref~\cite{CelataPRSTAB}. As already shown in this paper, the external magnetic 
field, by itself, alters the phase shift for a given cloud density. Given that the 
wiggler field is rather complex, along with a cloud density that is longitudinally nonuniform, 
simulations become particularly important to fully interpret results from such an 
experiment. In this paper, we examine just the effects of the nonuniform wiggler field 
on the phase shift.   

\begin{figure}[htb]
  \centering
  \includegraphics*[width=70mm]{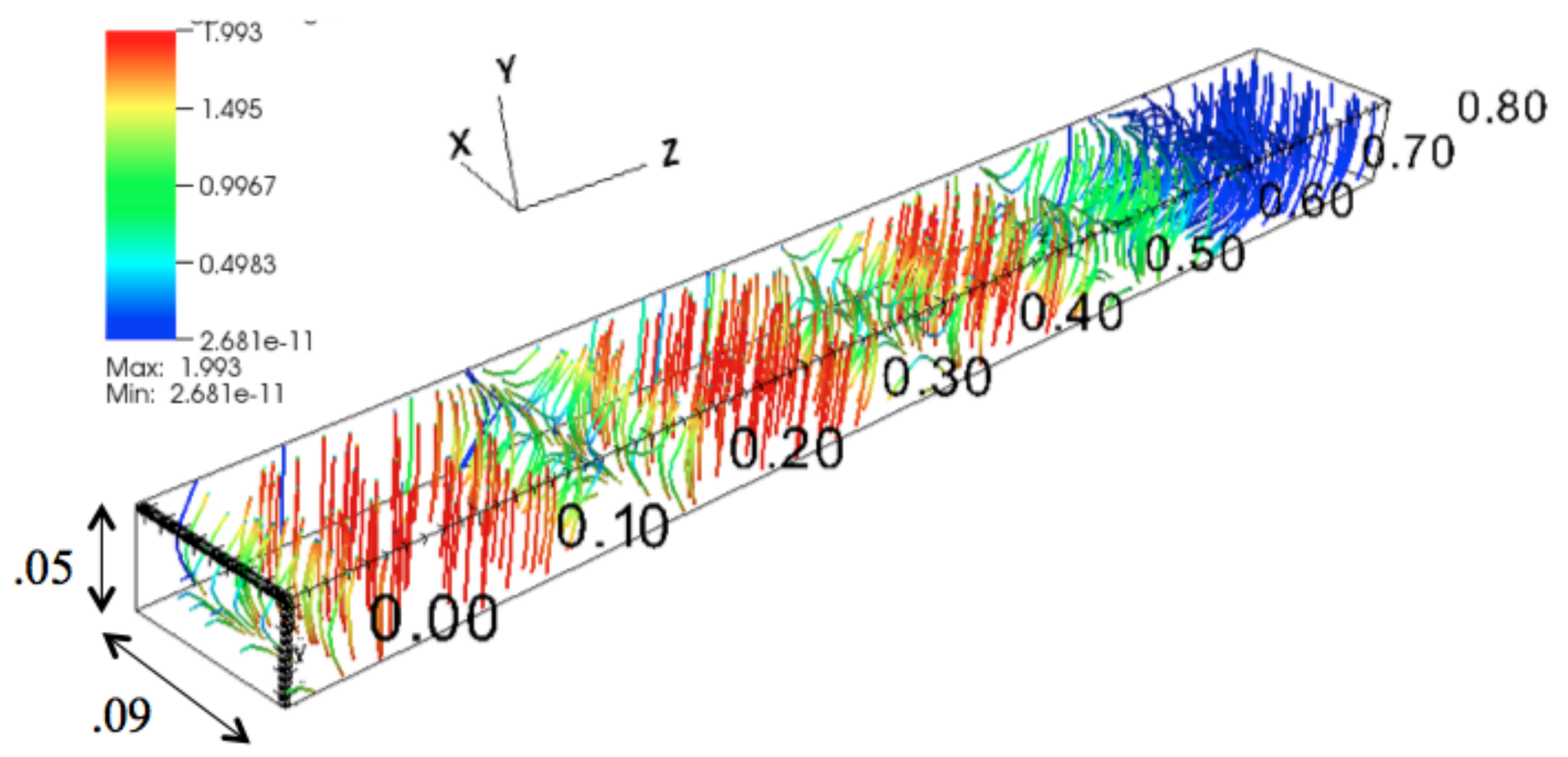}
  \caption{The full Wiggler field in 3 dimensions}
  \label{wigglerfield}
\end{figure}

The wiggler chamber cross section in CesrTA is close to that of a rectangle. The
height is 5cm and width is 9cm. The corners of the rectangle are chopped, so that 
the actual width of the base and top is 64.6mm and the height of the side walls are 
24.6mm. This which would only moderately alter the results obtained
from using a perfect rectangle. Thus, for the sake of simplicity, the simulations were
done with a perfect rectangle with the above parameters.  The length of the section 
simulated is 80cm, which corresponds to half the length of the wiggler. This is
sufficient to account all the variations in the wiggler magnetic field.  
The computed wiggler magnetic field used was based on the formulation given
in \cite{SaganPAC}. Figure~\ref{wigglerfield} shows the magnetic field, in
three dimensions. For most of the region, the field is oriented vertically, while in
the transition region between poles there is a longitudinal component to the field.
There is almost no magnetic field in the horizontal direction.      
\begin{figure}[htb]
 \centering
 \includegraphics*[width=70mm,height=40mm]{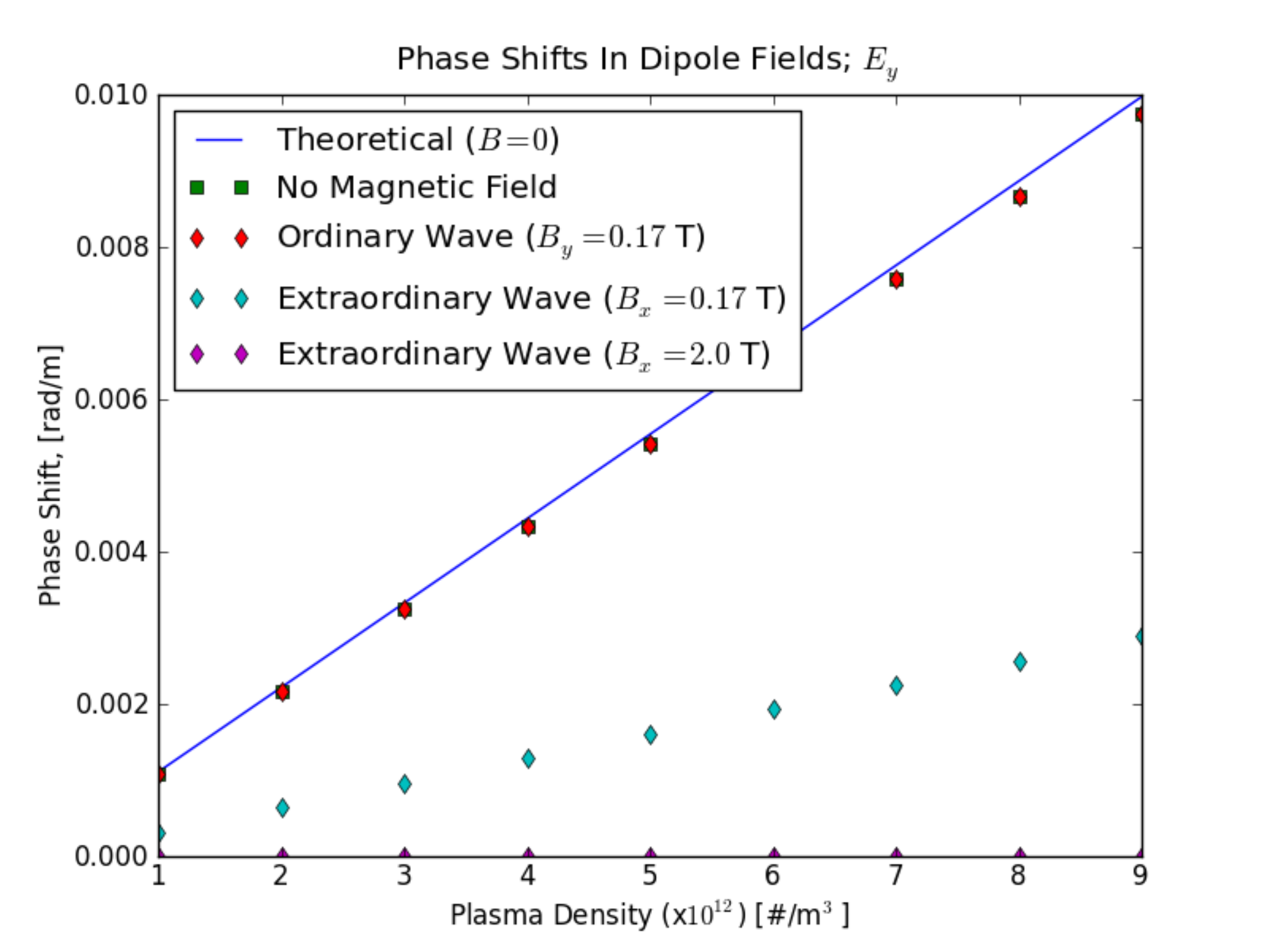}
 \caption{Variation of phase shift with cloud density for different
magnetic fields in a rectangular chamber with a vertical wave electric field.}
  \label{WiggDipVertWave}
\end{figure}
\begin{figure}[htb]
 \centering
 \includegraphics*[width=70mm,height=40mm]{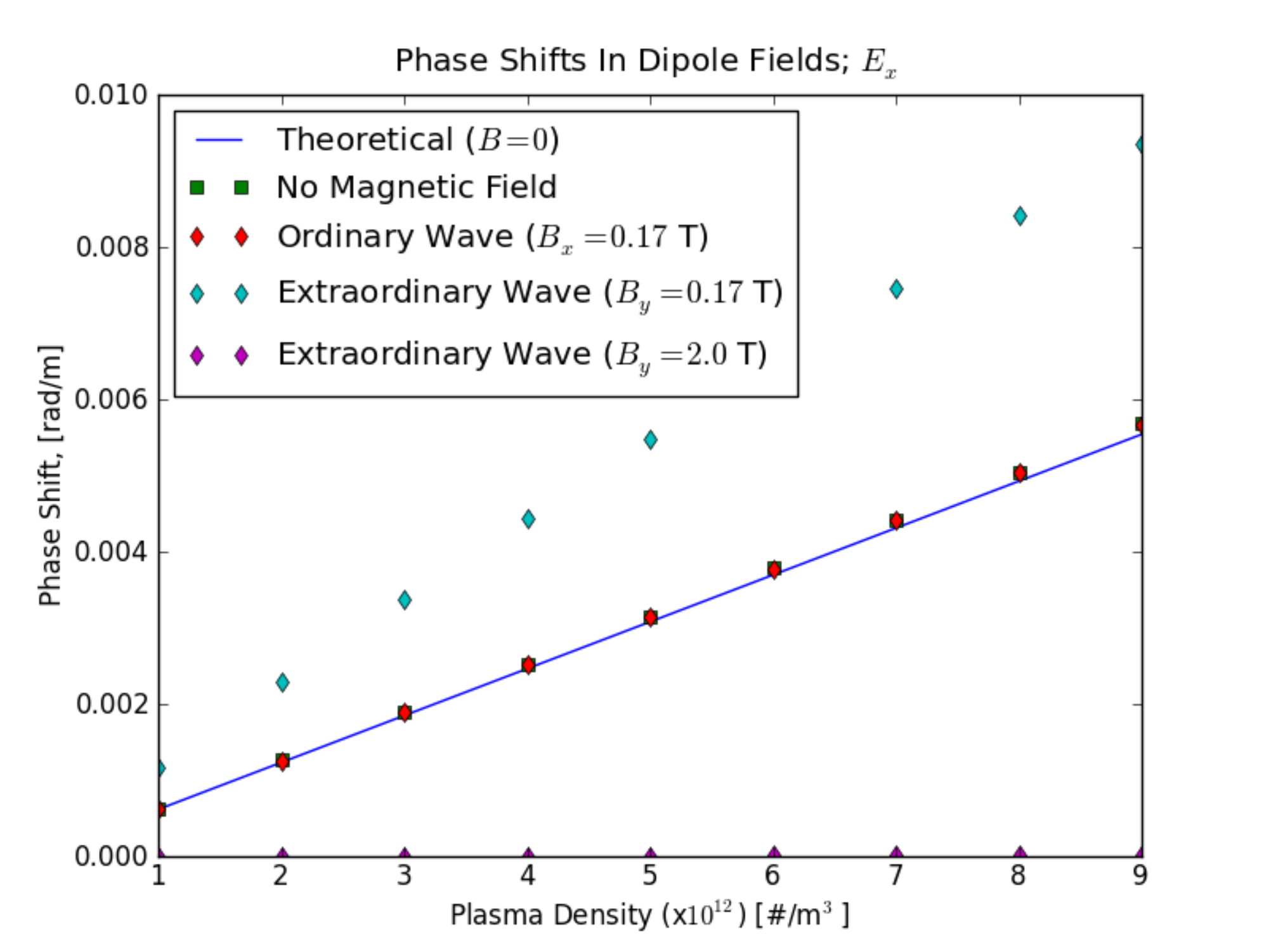}
 \caption{Variation of phase shift with cloud density for different
magnetic fields in a rectangular chamber with a horizontal wave electric field.}
  \label{WiggDipHorWave}
\end{figure}

Even before performing simulations with the full wiggler field turned on, 
some preliminary studies were done with a just a constant dipole field within
the same geometry. Since the shape of the vacuum chamber cross section is rectangular,
the the cutoff frequency of the wave is determined by the polarization of the
TE wave. If the wave electric field is pointing in the vertical direction,
the cutoff frequency is 1.66GHz and if the field is horizontal, the cutoff
is 3 GHz. All simulations were done at a frequency 10\% above the respective
cutoff. Figure \ref{WiggDipVertWave} shows the phase shift associated with
propagation of a vertically polarized wave under different conditions. The figure 
shows that when the wave electric field is polarized along the external 
magnetic field, the phase shift matches with that predicted by Eq~(\ref{phaseshift}). 
This result is expected to be true in the case of a rectangular 
cross section, where the wave electric field is pointing parallel to the external 
magnetic field everywhere in the pipe, and is thus unaffected by the external 
magnetic field. As already shown, this would not be true in the case of a curvature 
in the cross-section boundary. The figure also shows that the phase shift is 
suppressed in the case of the wave electric field pointing perpendicular to the external field,
referred to as extraordinary wave. The cyclotron resonance in this case occurs when the 
magnetic field is equal to $0.06$T and the field here was set to a much higher value.
In general, the wave electric field perturbs the electrons, causing them to oscillate and
thereby alter the wave dispersion relation. When the external magnetic field is very 
high, the electrons tend to get locked against any motion transverse to the  magnetic 
field. Since the wave electric field is perpendicular to the external magnetic field 
they will encounter electrons that tend to be "frozen". As a result the wave will 
undergo reduced electron cloud induced phase shift at magnetic fields much higher than 
that causing cyclotron resonance for the specific carrier frequency. 

Figure \ref{WiggDipHorWave} corresponds to a wave with the electric field
pointing horizontally. This shows similar features as those in Figure \ref{WiggDipVertWave}.
The wave frequency is 3.3GHz and cyclotron resonance occurs at field of 0.11 T, 
and so the figure shows enhanced phase shift at a magnetic field setting close to this value. 
Exciting the chamber at this higher frequency could be less efficient due
to poorer matching between the various hardware components. In addition,
one can expect a mixing of modes to take place at higher frequencies because of 
the presence of various irregularities in a real beam pipe as opposed to a simulated one. 
Nevertheless, studying this mode is important because the wiggler magnetic field is
mostly pointing in the vertical direction. One could amplify the signal by setting
a wiggler field so that a cyclotron resonance is excited. 
This effect might prove useful in detecting the presence of very low density
electrons in wiggler regions. The presence of low energy electrons in wiggler and undulators
is of particular interest if the device is cryogenic, in which case the electrons
are believed to contribute to the heat load of the system \cite{CasalbuoniPRSTAB}. 
The electron cloud in such systems would be produced by electron beams primarily through
photoemission, and thus if present, they will occur at very low densities, requiring greater 
sensitivity in the detection.  
\begin{figure}[htb]
  \centering
  \includegraphics*[width=70mm,height=40mm]{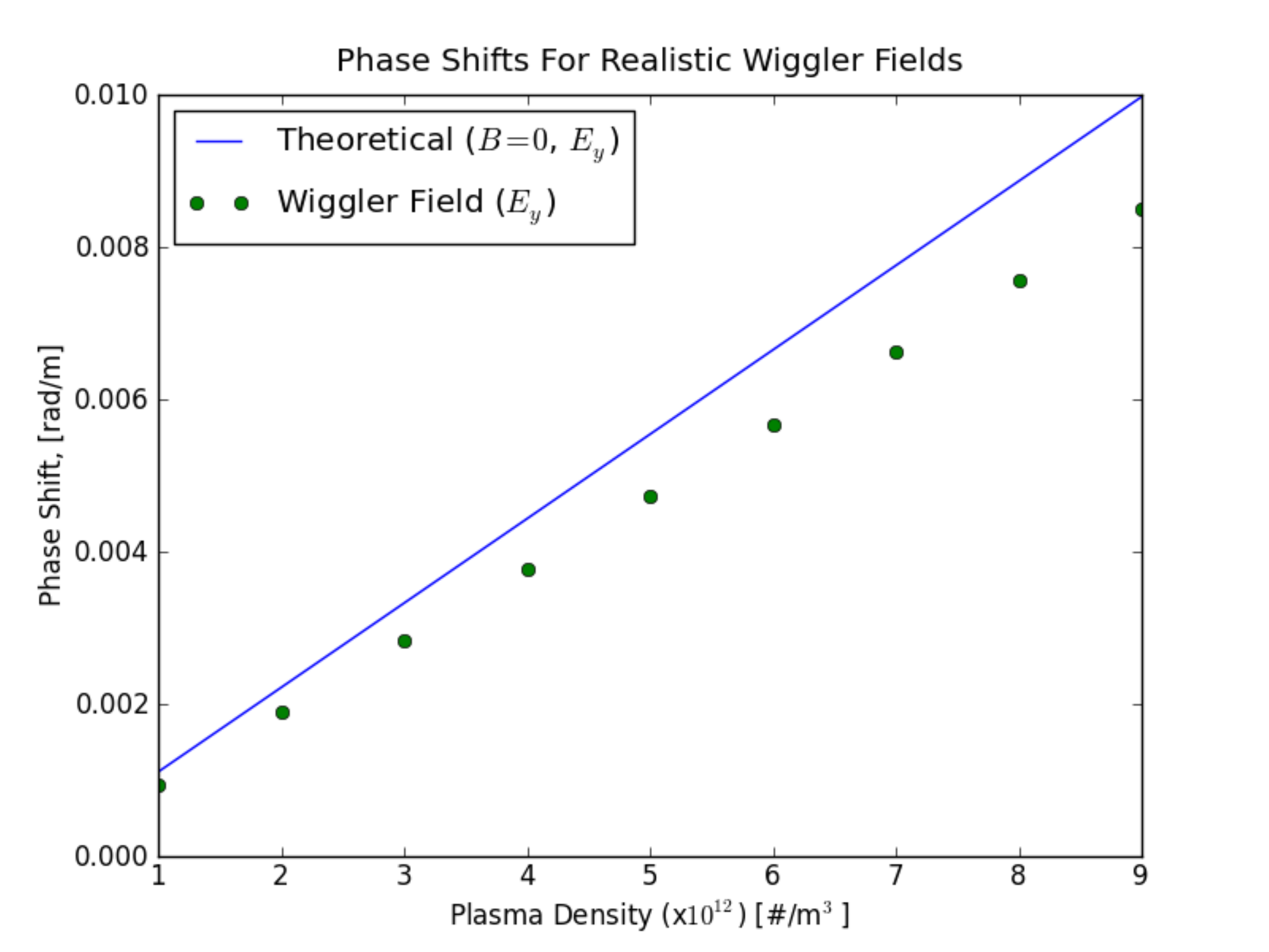}
  \caption{Variation of phase shift with cloud density with the 
  full wiggler magnetic field and a vertical wave electric field.}
  \label{WiggFullField}
\end{figure}

In the end, we look at the phase shift in the presence of the full wiggler field. 
Figure \ref{WiggFullField} shows that the phase shift gets suppressed by about 20\% 
when compared that expected in the absence of any fields. In this case, the wave electric 
field is pointing in the vertical direction, which is a configuration that should have little 
effect over the phase shift, since the external magnetic field is largely vertical. 
A longitudinal magnetic field would alter the dispersion relationship, where the wave gets 
split into a left and right circularly polarized components. This has been analyzed for guided wave 
propagation in cylindrical geometries in Ref~\cite{Uhm}.  While the 20\% reduction in phase 
shift in our result can be attributed to such an effect, a detailed analysis of the same is 
beyond the scope of this paper. Overall, it is clear that simulations are of prime importance 
to accurately interpret the observed electron cloud induced phase shifts across such wiggler fields. 

\section{Summary}
In this paper, we provide a comprehensive account of the simulation
and analysis effort that has been carried out in conjunction with the
experimental effort of using TE waves to measure electron clouds in
CesrTA. These simulations helped confirm several physical phenomena either
in a quantitative or in a qualitative manner. For example, they helped 
validate Eq~\ref{phaseshift} that relates the phase shift with cloud density 
for geometries like the CesrTA beam pipe, which does not have a regular 
shape such as rectangular or circular. The effect of reflections on phase shift
measurements has always been a concern, and simulations show that one must
be careful especially of standing waves excited within the beam pipe
due to partial reflectors.  The feasibility of using standing waves to
measure the cloud density is clearly demonstrated by simulations and has
provided valuable guidance to the experimental effort being carried out
at CesrTA.  In the process, we were able to determine a novel method of
detecting the presence of standing waves in simulations by averaging the 
total Poynting vector flux across a surface over time for varying frequencies.  

Simulations of phase shifts in the presence of external magnetic fields were 
modeled for a variety of cases. The nature of results vary greatly based on the
the parameters present in the system. For example, the possibility of
exciting cyclotron resonances is clearly shown in simulations, which would
be possible to produce in dipole fields present in a chicane. However,
dipole fields used in bend regions of an accelerator are much higher
and they can suppress the electron induced phase shift, depending on the
polarization of the wave. In the presence of curved boundaries, there is
always a mixture of effects from ordinary and extraordinary wave
propagation. Since most accelerator vacuum chambers have a curvature,
this effect is important to understand. A direct comparison with the
analytic expression Eq~\ref{disprel} would lead to an incorrect interpretation 
of the measured data. 

The results will always be hard to interpret when a
system is near the cyclotron resonance, when the phase shift is theoretically
infinity. Simulations and experiments would never yield an infinity,
and there may be poor agreement between the two in such regimes. However,
an enhancement of the signal that is still predictable can always be obtained 
by moving reasonably close to a cyclotron resonance point. Electron cloud 
formation in wiggler fields can be very important in machines such as positron 
damping rings. Given the complexity of such a system, experiments of determining
the cloud density using TE waves have to be accompanied by careful simulations 
before interpreting any results obtained from measurements.

It may be noted that the relationship between the phase shift and
cloud density is always linear regardless of how complex the system
is. This would be true for very low cloud densities, which would have
low plasma frequencies. Thus, this method can be of great utility if
one is interested in relative changes in electron cloud densities for
example when a machine is undergoing conditioning. Simulations can then
be used to obtain a proportionality constant between cloud density and
phase shift.

The studies in this paper were always done with an electron density
that was cold and uniformly distributed transversely and longitudinally. 
Simulations have not indicated a dependence on temperatures associated 
with typical electron cloud densities. In a transverse nonuniform distribution, there 
will be a slight enhancement in the phase shift if more electrons are populated in
regions with high peak electric fields produced by the wave. As mentioned
earlier, longitudinal variation of the cloud density becomes important
in wiggler fields because it couples with the longitudinal variation
of the magnetic field. These additional complexities could be topics for
future studies.

The TE wave method is an attractive technique for measuring electron
cloud densities that can replace or complement other measurement
methods. Besides CesrTA, this method is being studied at other accelerator 
facilities \cite{Friedman,Eddy,Thangaraj}.
The measurement technique and its required instrumentation are simple,
the process is noninvasive, and can be kept in operation continuously.
Thus, it holds the promise of wide usage wherever it is useful to
monitor the electron cloud properties continuously and at all locations
of the accelerator. It is evident that a careful study toward understanding
of the physical phenomenon through analysis and simulation are
very important toward proper interpretation of the measured data.

\section{Appendix: Derivation of Dispersion Relationship} 
In this Appendix, we provide a derivation of Eq~\ref{disprel} 
which is not specific to the geometry of the cross-section of
the waveguide. We also discuss the approximations and assumptions 
associated with the derivation of this equation. This dispersion 
relationship is specific to guided waves propagating through electron 
clouds in field free regions. The starting equations for such a system 
would include the fluid and Maxwell's equations. These are, 
\begin{eqnarray}
 & & m[ \frac{\partial {\bf v}}{\partial t} + ({\bf v}\cdot\nabla){\bf v}] +  
 e({\bf E} + {\bf v} \times {\bf B})  = 0 \nonumber \\
 & &  \frac{\partial n_e}{\partial t} + \nabla \cdot (n_e{\bf v}) = 0 \nonumber \\
 & & \nabla \cdot {\bf E} = \frac{en_e}{\epsilon_0} \nonumber \\ 
 & & \nabla \cdot {\bf B} = 0 \nonumber \\ 
 & & \nabla \times {\bf E} = - \frac{\partial {\bf B}}{\partial t} \nonumber \\
 & & \nabla \times {\bf B} = \mu_0(-en_e{\bf v} + {\bf J_{ext}}) + \mu_0 \epsilon_0 
\frac{\partial {\bf E}}{\partial t}
\label{unperturbeqs}
\end{eqnarray}
where $n_e$ is the number density of the electrons, ${\bf v}$ is the velocity of the
fluid, and ${\bf J_{ext}}$ is an external current density. The other terms have their 
usual meanings. The continuity equation is not independent from the rest of the equations
as it can be obtained from Maxwell's equations.

We perturb all quantities about an equilibrium, so that we have
${\bf v} = {\bf v_0} + {\bf v^{(1)}}$, $n_e = n_0 + n^{(1)}$, ${\bf E} = {\bf E_0} + {\bf E^{(1)}}$,
${\bf B} = {\bf B_0} + {\bf B^{(1)}}$
where the zeroth order quantities satisfy the steady state condition
${\partial}/{\partial t} = 0$. As a result, we get
 \begin{eqnarray}
 & & m({\bf v_0}\cdot\nabla){\bf v_0} +
 e({\bf E_0} + {\bf v_0} \times {\bf B_0}) = 0 \nonumber \\
 & & \nabla \cdot (n_0{\bf v_0}) = 0 \nonumber \\
 & & \nabla \cdot {\bf E_0} = \frac{en_0}{\epsilon_0} \nonumber \\ 
 & & \nabla \cdot {\bf B_0} = 0 \nonumber \\ 
 & & \nabla \times {\bf E_0} = 0 \nonumber \\ 
 & & \nabla \times {\bf B_0} = \mu_0(-en_0{\bf v_0} + {\bf J_{ext}}) 
\label{equilibeqs} 
\end{eqnarray}

Such an equilibrium state requires for the particles to be confined
in a steady state indefinitely. This is normally associated with neutral plasmas
in which the ions may be considered immobile, or charged particles trapped for a 
long period of time in confinement devices. In this paper, the
electron cloud density is sustained for the duration of the bunch train passage.
The above equilibrium condition may be considered valid as long as $1/\tau \gg f$ 
where $\tau$ is time of confinement of the charge and $f$ is the frequency of
the perturbing wave. Periodic changes in the state occurring over time scales greater 
than the wave periodicity manifest themselves as modulations of the output signal, while 
changes occurring over much smaller time scales, remain unresolved by the carrier wave. 
Thus, the spectrum of the output signal would depend on the variation of the electron 
cloud associated with its build up and decay.  
 
Inserting the perturbation expansion into the original fluid and Maxwell's equation,
and imposing the above equilibrium conditions and ignoring terms of second and higher
order, we get,
\begin{eqnarray}
 & & m{\frac {\partial {\bf v^{(1)}}}{\partial t}} + e({\bf E^{(1)}} + {\bf v^{(1)}} \times {\bf B_0}) = 0\nonumber \\ 
 & &  \frac{\partial n^{(1)}}{\partial t} + \nabla \cdot (n_0{\bf v^{(1)}}) =  0 \nonumber \\
 & & \nabla \cdot {\bf E^{(1)}} = \frac{en^{(1)}}{\epsilon_0} \nonumber \\ 
 & & \nabla \cdot {\bf B^{(1)}} = 0 \nonumber \\ 
 & &  \nabla \times {\bf E^{(1)}} = -{\bf \frac {\partial B^{(1)}}{\partial t}} \nonumber \\
 & & \nabla \times {\bf B^{(1)}} = \mu_0(-en_0{\bf v^{(1)}} +  
  \epsilon_0 \frac{\partial{\bf  E^{(1)}}}{\partial t}) 
\label{perturbeqns}
\end{eqnarray}

These equations are linear and we can seek all perturbations to have the following form,
\begin{eqnarray}
  {\bf v}^{(1)}(x,y,z,t) = \tilde{{\bf v}}(x,y)e^{i(k - \omega t)} \nonumber \\
  n^{(1)}(x,y,z,t) = \tilde{n}(x,y)e^{i(k - \omega t)} \nonumber \\
  {\bf E}^{(1)}(x,y,z,t) = \tilde{{\bf E}}(x,y)e^{i(k - \omega t)} \nonumber \\
  {\bf B}^{(1)}(x,y,z,t) = \tilde{{\bf B}}(x,y)e^{i(k - \omega t)} 
\label{wavesoln}
\end{eqnarray}
From the above, it is clear that
\begin{equation}
  \frac{\partial}{\partial z} = ik, ~~~~~~~ 
  \frac{\partial}{\partial t} = -i\omega
\label{waveform}
\end{equation}

This form is valid as long as the geometry along ``$z$", the longitudinal
coordinate is uniform and infinite. This is not entirely true when
there are partial reflectors. In the case of perfect reflectors, one
would obtain discrete values for $k$, representing standing waves.
Thus one can expect that the above form to be more accurate when
close to a resonance in the presence of partial reflectors. In general,
the reflectors would be small enough so that the above form of
solutions can be considered a valid approximation.  

For the sake of convenience, we drop the accent ~$\tilde{}$~ and the arguments
$(x,y)$ in the functions given in Eq~(\ref{wavesoln}). 
To simplify the analysis, we make two assumptions, (1) The fluid is cold and at rest. 
So ${\bf v_0} = 0$, and (2) the density is uniform, which means $n_0$ = constant.
We further assume that there is no external magnetic field, which means that
${\bf B_0} = 0$. In the absence of any static external magnetic fields, the
only contribution to ${\bf B_0}$ would be the magnetic field produced by the beam.
Since the beam is highly relativistic, this would be confined along the length of
the bunch. It is reasonable to disregard this when the gap between the bunches
is much larger than the bunch length, in which case the wave would sample mostly
a field free region.      

Inserting the waveform solutions (Eq~\ref{wavesoln}) into the perturbed
momentum and continuity equations in Eq~(\ref{perturbeqns}), with
${\bf B_0} = 0$ and using the relationships of Eq~(\ref{waveform}), we have, 
\begin{eqnarray}
& & -im\omega {\bf v^{(1)}} + e{\bf E^{(1)}} = 0 \nonumber \\
& & -i\omega n^{(1)} + n_0 \nabla \cdot {\bf v^{(1)}} = 0 
\label{fluidwaveeq}
\end{eqnarray}
Combining these, we get 
\begin{equation}
 n^{(1)} + \frac{n_0e}{m\omega^2} \nabla \cdot {\bf E^{(1)}} = 0. 
\end{equation}
Combining this with the perturbed electrostatic field equation  in Eq~(\ref{perturbeqns})
gives $n^{(1)} = 0$, which means, up to the first order, there is no perturbation in the
charge density due to the wave electric and magnetic fields. This gives us 
\begin{equation}
\nabla \cdot {\bf E^{(1)}} = 0
\label{wavegauss}
\end{equation}
implying that the wave is purely electromagnetic.
By combining the first (momentum), and the last (Ampere's law) equation of  
Eq~(\ref{perturbeqns}), and using the relationships of Eq~(\ref{waveform}), 
we get after some algebra,
\begin{equation}
\nabla \times {\bf B^{(1)}} = -i\omega\mu_0\epsilon {\bf E^{(1)}} 
\label{waveampere}
\end{equation}
where $\epsilon = \epsilon_0(1 - \omega_p^2/\omega^2)$. Similarly,
it is easy to see that,
\begin{equation}
\nabla \times {\bf E^{(1)}} = i\omega{\bf B^{(1)}}
\label{wavefaraday}
\end{equation} 

Using Eqs~\ref{wavegauss}, \ref{waveampere} and \ref{wavefaraday}, along with
$\nabla \cdot {\bf B^{(1)}} = 0$, and assuming that the boundary conditions
are perfectly conducting, one can follow the steps given in Ref\cite{Jackson}.
to get
\begin{equation}
 \gamma^2 = \mu_0\epsilon\omega^2 - k^2.
\label{eigenwaveeq}
\end{equation}
The constant $\gamma^2$ must be nonnegative for oscillatory solutions, and will 
take on a set of discrete ``eigenvalues", corresponding to the different modes 
associated with the geometry of the cross-section of the waveguide. Combining
Eq~(\ref{eigenwaveeq}) with a similar relationship for a vacuum waveguide where,
$\epsilon = \epsilon_0$, one can easily show that,
\begin{equation}
k^2 = \frac{\omega^2}{c^2} - \frac{\omega_p^2}{c^2} - \frac{\omega_{co}^2}{c^2}
\end{equation}
$\omega_{co} = c\gamma$ being the angular cutoff frequency for the vacuum waveguide. 
This relationship is the same as Eq~(\ref{disprel}). 

\section*{Acknowledgements}
The authors wish to thank John Sikora for many useful discussions
and for suggesting us to do the simulations with partial internal 
reflections. Thanks to Jim Crittenden for helping us in 
generating the complete the wiggler magnetic field. We also wish 
to thank David Rubin, Mark Palmer and Peter Stoltz for their support 
and guidance. This work was supported 
by the US National Science Foundation (PHY-0734867, PHY-1002467, 
and PHY-1068662) and the US Department of Energy (DE-FC02-08ER41538
and DE-SC0006505; DE-FC02-07ER41499 as part of the ComPASS SCiDAC-2 project,
and DE-SC0008920 as part of the ComPASS SCiDAC-3 project).

\bibliographystyle{model1a-num-names}

\begin{thebibliography}{24}
\bibitem{Caspers}
F. Caspers, W. Hofle, J. M. Jimenez, J. F. Malo, J. Tuckmantel, and T. Kroyer, 
in Proceedings of the 31st ICFA Beam Dynamics Workshop: Electron Cloud Effects
(ECLOUD04), Napa, California 2004 (CERN Report No. CERN-2005-001, 2004).
\bibitem{Kroyer}
T. Kroyer , F. Caspers, E. Mahner, Proceedings of 2005 Particle Accelerator Conference, 
Knoxville, Tennessee 2212-2214
\bibitem{StefanoPRL}
S. De Santis, J. M. Byrd, F. Caspers, A. Krasnykh, T. Kroyer, M. T. F. Pivi, and K. G. Sonnad
Phys. Rev. Lett. 100, 094801 (2008)
\bibitem{SonnadPAC}
Kiran Sonnad, Miguel Furman, Seth Veitzer, Peter Stoltz and John Cary,
Proceedings of PAC07, Albuquerque, New Mexico, USA, pp. THPAS008
\bibitem{SonnadANKA} 
K G Sonnad {\it et al}
Proceedings of Particle Accelerator Conference 2009, Vancouver, Canada, 2009, pp.TH5RFP044
\bibitem{SikoraIPAC11}
J.P. Sikora {\it et al}, Proceedings of International Particle Accelerator Conference 2011, San Sebastián, Spain, pp.TUPC170
\bibitem{SonnadAPS}
K.G. Sonnad {\it et al} http://meetings.aps.org/link/BAPS.2007.DPP.TP8.134
49th Annual Meeting of the Division of Plasma Physics
\bibitem{VeitzerDOE}
S. Veitzer, DOE Scientific and Technical Information,  http://www.osti.gov/bridge
Identifier number 964651
\bibitem{PiviEPAC}
M. T. F. Pivi, {\it et al}, Proceedings of European Particle Accelerator Conference 2008, Genoa, Italy
pp. MOPP065
\bibitem{Vorpal}
C. Nieter and J. R. Cary, J. Comp. Phys. 196, 448-472 (2004).
\bibitem{Uhm}
H. S. Uhm, K. T. Nguyen, R. F. Schneider an d J. R. Smith,
Journal of Applied Physics, Vol 64(3), 1988, pages 1108-
1115.
\bibitem{Berenger}
J. Berenger,  Journal of Computational Physics 114, 185 (1994)
\bibitem{Yee}
K. Yee,   IEEE Transactions on Antennas and Propagation, AP-14, 302 (1966)
\bibitem{DeSantisPRSTAB}
S. De Santis, Phys. Rev. ST Accel. Beams 13, 071002 (2010)
\bibitem{SikoraNIM}
John Sikora and Stefano DeSantis, http://arxiv.org/abs/1311.5633  
\bibitem{DeSantisPAC11}
S. De Santis {\it et al}, Proceedings of 2011 Particle Accelerator Conference, New York, NY, USA,
pp. MOP228
\bibitem{GoldstoneRutherford}
R J Goldstone and P H Rutherford, Introduction to Plasma Physics
Institute of Physics Publishing, 1995
\bibitem{CelataPRSTAB}
C. Celata Phys. Rev. ST - Accel. Beams 14, 041003 (2011)
\bibitem{SaganPAC}
D. Sagan, J. A. Crittenden, D. Rubin and E. Forest, Proceedings of Particle Accelerator Conference 2003, Portland, OR, USA pp.1023
\bibitem{CasalbuoniPRSTAB}
S. Casalbuoni, S. Schleede, D. Saez de Jauregui, M. Hagelstein, and P. F. Tavares,
Phys. Rev. ST Accel. Beams 13, 073201 (2010) 
\bibitem{Friedman}
S. Federmann, F. Caspers, and E. Mahner, Phys. Rev. ST Accel. Beams 14, 012802 (2011)
\bibitem{Eddy}
N. Eddy, J. Crisp, I. Kourbanis, K. Seiya, B. Zwaska, S. De Santis
Proceedings of Particle Accelerator Conference 2009, Vancouver, BC, Canada pp. WE4GRC02
\bibitem{Thangaraj}
J. C. Thangaraj, N. Eddy, B. Zwaska, J. Crisp, I. Kourbanis, K. Seiya
Proceedings of the Electron Cloud Workshop 2010, Ithaca, New York, USA pp. DIA00
\bibitem{Jackson}
J D Jackson. Classical Electrodynamics, Wiley, New York, NY, 3rd ed. edition, (1999)
\end{thebibliography}

\end{document}